\documentclass[10pt,journal,TCSS]{IEEEtran}
\usepackage{amsmath,amsfonts}
\usepackage{amssymb}
\usepackage{stfloats}
\usepackage{multirow}
\usepackage{booktabs}

\usepackage{makecell}
\usepackage{color}
\usepackage{float}
\usepackage{algorithm}
\usepackage{algorithmic}

\usepackage{cite}

\usepackage[caption=false,font=normalsize,labelfont=sf,textfont=sf]{subfig}
\usepackage{graphicx}
\graphicspath{{./figs/}}
\usepackage{array}
\usepackage{textcomp}

\usepackage{url}
\usepackage{verbatim}

\usepackage{balance}
\usepackage{enumerate}

\begin{document}

\title{Node Injection Attack Based on Label Propagation Against Graph Neural Network}
\author{Peican Zhu,~\IEEEmembership{Member,~IEEE, } Zechen Pan, Keke Tang,~\IEEEmembership{Member,~IEEE, } Xiaodong Cui, Jinhuan Wang, Qi Xuan,~\IEEEmembership{Senior Member,~IEEE}

\thanks{This work has been submitted to the IEEE for possible publication. Copyright may be transferred without notice, after which this version may no longer be accessible.

This work was supported in part by the Key R\&D Program of Zhejiang (Grant no. 2022C01018), National Natural Science Foundation of China (Grant nos. 62073263, U21B2001, 62102105), Fundamental Research Funds for the Central Universities (Grant no. D5000230112), Open Research Subject of State Key Laboratory of Intelligent Game (Grant no. ZBKF-24-02).
\emph{(Corresponding author: Xiaodong Cui and Qi Xuan.)}}

\thanks{P. Zhu is with the School of Artificial Intelligence, Optics and Electronics (iOPEN), Northwestern Polytechnical University (NWPU), Xi’an 710072, Shaanxi, China (e-mail: ericcan@nwpu.edu.cn).}

\thanks{Z. Pan is with the School of Computer Science, NWPU, Xi’an 710072, Shaanxi, China (e-mail: 928598047@mail.nwpu.edu.cn).}

\thanks{K. Tang is with the Cyberspace Institute of Advanced Technology, Guangzhou University, Guangzhou 510006, Guangdong, China (e-mail: tangbohutbh@gmail.com).}
\thanks{X. Cui is with the School of Marine Science and Technology, NWPU, Xi’an 710072, Shannxi, China (e-mail: xiaodong.cui@nwpu.edu.cn).}
\thanks{J. Wang is with the Institute of Cyberspace Security, College of Information Engineering, Zhejiang University of Technology, Hangzhou 310023, Zhejiang, China (e-mail: jhwang@zjut.edu.cn).}
\thanks{Q. Xuan is with the Institute of Cyberspace Security, College of Information Engineering, Zhejiang University of Technology, Hangzhou 310023, Zhejiang, China, and also with the Binjiang Institute of Artificial Intelligence, Zhejiang University of Technology, Hangzhou 310056, Zhejiang, China (e-mail: xuanqi@zjut.edu.cn).}}

\maketitle
\begin{abstract}
Graph Neural Network (GNN) has achieved remarkable success in various graph learning tasks, such as node classification, link prediction and graph classification. The key to the success of GNN lies in its effective structure information representation through neighboring aggregation. However, the attacker can easily perturb the aggregation process through injecting fake nodes, which reveals that GNN is vulnerable to the graph injection attack. Existing graph injection attack methods primarily focus on damaging the classical feature aggregation process while overlooking the neighborhood aggregation process via label propagation. To bridge this gap, we propose the label-propagation-based global injection attack (LPGIA) which conducts the graph injection attack on the node classification task. Specifically, we analyze the aggregation process from the perspective of label propagation and transform the graph injection attack problem into a global injection label specificity attack problem. To solve this problem, LPGIA utilizes a label propagation-based strategy to optimize the combinations of the nodes connected to the injected node. Then, LPGIA leverages the feature mapping to generate malicious features for injected nodes. In extensive experiments against representative GNNs, LPGIA outperforms the previous best-performing injection attack method in various datasets, demonstrating its superiority and transferability.
\end{abstract}

\section{Introduction}
\IEEEPARstart{W}{ith} the adoption of the message passing scheme, Graph Neural Network (GNN) is capable of handling graph data through aggregating the structure information. Due to this special scheme, GNN has achieved remarkable success in various fields including node classification \cite{gcn2017,gat2018,fagnn2022,ginsd2024,kyc2024}, link prediction \cite{sage2017,compgcn2020,dgcn2023}, graph classification \cite{gin2019,ssngc2021}, recommender system \cite{ahgrr2023} and emotion-cause pair extraction \cite{ecpe2024}.  However, recent studies have revealed that deep learning models can be easily fooled by adversarial inputs \cite{uav2022,ham2024,lesson2024,gafsi2024}. In this case, the message passing scheme in GNN fails to identify the fake message propagating in the graph, leading to concerns about the robustness of GNNs \cite{nettack2018, rls2v2018}.

Following pioneering research, many scholars have devoted their efforts to developing adversarial attack methods in order to analyze the robustness of GNN. In \cite{mettack2019,pgd2019,linkatk2020,sga2021,graphatker2022,gf2023}, the authors proposed to pass wrong messages through modifying the original graph structure, which are referred to as Graph Modification Attack (GMA). Nevertheless, it is usually unrealistic for the attacker to possess the authority to modify the original links or features of nodes. For instance, in social networks, it is difficult for attackers to modify the personal information of the target user. However, it is relatively easy for the attacker to create a new account that conducts interaction with the target user. Thus, to address the drawback of GMA, Sun \emph{et al.} \cite{nipa2020} proposed Graph Injection Attack (GIA) which is a new category of adversarial attack.

For GIA, the attack is performed through injecting fake nodes which carry malicious features into the original graph. Hence, wrong messages can successfully propagate without modifying the existing links and features. In contrast to GMA, GIA encounters unique and essential challenges, among which is the key problem of connecting the injected nodes with the original nodes to disrupt the message passing process of GNN. Wang \emph{et al.} \cite{afgsm2020} used approximate gradients of the loss function to guide the connection process. However, such approximation fails to explicitly explore the graph structure information. Thus, through investigating the feature aggregation process of GNN, Zou \emph{et al.} \cite{tdgia2021} proposed to connect the injected nodes with the node possessing higher topological vulnerability which is related to the degree of the node. Additionally, to flexibly model the feature aggregation process, Tao \emph{et al.} \cite{gnia2021} optimized the connections through training a neural network. From a different perspective, Wang \emph{et al.} \cite{ca2022} successfully transformed the GIA into a graph clustering problem. They defined an adversarial feature for each victim node respectively and then clustered nodes accordingly. Overall, existing GIA methods primarily design connection strategies based on destroying the feature aggregation process of GNN. The key to those attacks is that the injected node propagates malicious features to the original nodes through the classical feature aggregation process of GNN. However, those GIA methods overlook the neighborhood aggregation process via label propagation in GNN. Compared with classical feature aggregation, neighborhood aggregation via label propagation can better leverage the node prediction information and the graph structure information \cite{appna2019,gcnlpa2021,pta2021,cs2021}. As existing GIA methods mainly consider the feature aggregation process, they fail to enhance their attack performance through effectively leveraging node prediction information. To better utilize node prediction information instead of solely relying on the information from the feature aggregation process, efforts still need to be spared to design a GIA method considering the message passing scheme via node prediction.

To bridge this gap, inspired by the idea of label propagation \cite{lp2002,llg2003,lpch2007,cs2021}, we propose the label-propagation-based global injection attack (LPGIA). In order to apply label propagation into GIA scenarios, we reinterpret the objective of the GIA and transform the GIA problem into a label specificity attack issue \cite{lsa2022,g2snia2023}. Compared with the original label specificity attack problem, LPGIA mainly focuses on the global injection attack scenario instead of attacking a single node. Specifically, the target label is defined for each node in the original graph respectively. Accordingly, we aim to increase the probability of all nodes being classified as their target label, thereby inducing erroneous prediction. To solve the problem, LPGIA consists of two parts, i.e., cluster derivation and feature generation. In the cluster derivation procedure, we design a label propagation-based method to select the optimal combination of the victim nodes for each injected node. Then, in the feature generation procedure, our method generates the malicious features for injected nodes based on feature mapping. The performance of LPGIA is evaluated on representative datasets against five well-known GNNs on the node classification task. Through utilizing the idea of label propagation, LPGIA exhibits excellent performance, demonstrating its feasibility.

Overall, the contributions of this manuscript are listed as:

\begin{enumerate}[(1)]
\item We analyze GIA from the perspective of label propagation and transform the GIA problem into a global injection label specificity attack problem. Specifically, we define a target label for each node in the original graph respectively. With this knowledge, we reinterpret the objective of GIA to adapt label propagation to our scenarios.
\item We propose LPGIA, a global graph injection attack method based on label propagation, which aims to mislead GNN to generate erroneous predictions and thereby reduce its performance on the node classification task. To achieve the objective, LPGIA finds the optimal connection strategy from the perspective of label propagation and designs the malicious features according to the feature mapping.
\item We conduct extensive experiments against different kinds of GNN to evaluate the performance of our method. Compared with other baselines, the experimental results show that our proposed approach significantly deteriorates the prediction accuracy of GNN. Further, we also analyze the contribution of each module in our method and explore alternative strategies.

\end{enumerate}

The rest of this paper is organized as follows. In Section \ref{sec_rw}, the related works are provided in detail. Then, Section \ref{sec_pre} clarifies the definitions of GNN and GIA. In Section \ref{sec_method}, we introduce our motivation and propose our attack method. Later, sufficient experimental results are presented in Section \ref{sec_exp}. Finally, we conclude our work and discuss the future directions in Section \ref{sec_con}.

\section{Related Works}\label{sec_rw}
In this section, we briefly review some recent works related to the adversarial attack on GNNs, which could be generally categorized into two primary types, i.e., GMA and GIA. During the early stage, works on adversarial attacks against GNN focused on modifying the original graph, i.e., edges or node features. Dai \emph{et al.} \cite{rls2v2018} introduced the reinforcement learning technique to the targeted attack which aims to change the prediction of the target node. However, their method solely focused on edge modifying, which can not be used to modify features. Z\"{u}gner \emph{et al.} \cite{nettack2018} conducted the targeted attack by greedy searching which took into account both edge perturbations and feature perturbations to identify the optimal perturbation. In addition to the targeted attack, Z{\"u}gner and G{\"u}nnemann \cite{mettack2019} conducted the global attack in order to worsen the accuracy of GNN. They proposed a meta gradient-based strategy for the iterative selection of the most influential edge for the attack. Following such pioneering works, a number of improved methods based on gradient information emerged \cite{pgd2019,sga2021,mga2021,epoatk2023}.

The above methods belonging to the GMA category could not efficiently address the attacker's capability to modify the original data, due to the expensive cost \cite{cana2023}. To address this concern, Wang \emph{et al.} \cite{afgsm2020} extended the gradient-based method to the GIA scenario which only injected fake nodes into the graph. To fully leverage the advantages of GIA, Zou \emph{et al.} \cite{tdgia2021} proposed a novel framework that adopted a heuristic strategy to select the defective nodes for the injection based on their analysis of the feature aggregation process of GNN. However, the proposed weight highly relied on the node's degree which limited the ability to further explore graph structure information. Then, Tao \emph{et al.} \cite{gnia2021} developed a flexible injection attack approach through training a neural network to adaptively model the influence of node injection. To deceive the pruning mechanism in GNN, Chen \emph{et al.} \cite{hao2022} proposed a method to increase the feature similarity between the injected node and its neighbors. Furthermore, they argued that the destructive power of GIA might come from the damage to the homophily of the original graph. Along with this idea, Fang \emph{et al.} \cite{gani2022} proposed an evolutionary algorithm-based method to find the optimal combination of nodes and used the decrease of node homophily as a sorting index. From a different perspective that the GIA problem could be reformulated as the graph clustering problem, Wang \emph{et al.} \cite{ca2022} solved this problem based on Euclid’s distance between victim nodes’ adversarial feature vectors. Ju \emph{et al.} \cite{g2a2c2023} formulate GIA as a Markov decision process and leverage the rewards calculated from the model feedback to guide their attacks.

In summary, those GIA methods mainly consider the effect of the attack on the feature aggregation process and rely on the gradient information from the surrogate model. Since the importance of node features varies among different GNNs, the transferability of those methods will decrease when the attacker has no knowledge about the defense model. Unlike most previous works \cite{afgsm2020,tdgia2021,ca2022}, we leverage the node prediction information to conduct the attack, which is inherently related to the attacker's goal. Similarly, Tao \emph{et al.} \cite{gnia2021} use node prediction as input to guide the direction of prediction change. However, they still focus on the feature aggregation process and ignore the latent association between the node label and graph structure. Hence, we analyze the label propagation process to better utilize the node prediction information and graph structure information, resulting in remarkable attack performance improvements.

\section{Preliminaries}\label{sec_pre}
In this section, we first introduce some fundamentals related to GNN which serves as the surrogate model. Then, the details of GIA are clarified.

\subsection{Graph Neural Network}
Let $G=(\mathcal{V}, \mathcal{E})$ denote an undirected and unweighted graph, where $\mathcal{V}$ indicates the node set with a size of $n$ and $\mathcal{E}$ denotes the edge set. We assume that $A\in \{0,1\}^{n\times n}$ represents the symmetric adjacency matrix, where $A_{uv}=1$ if $(u, v)\in \mathcal{E}$, otherwise $A_{uv}=0$. Let $X\in \mathbb{R}^{n\times m}$ represent the corresponding feature matrix, where $m$ indicates the dimension of the feature vector.

For the node classification task, each node has a ground-truth label $c \in l=\{l_1, \dots, l_L\}$, where $L$ is the number of the categories in the label set $l$. Then GNN aims to derive the prediction of unlabeled nodes leveraging $A$ and $X$. As in \cite{gcn2017}, a classical structure of the GNN can be described as:
\begin{equation}
Z = f_\theta (A, X)=softmax(\tilde{A}\sigma (\tilde{A}XW_1)W_2),
\end{equation}
where $\tilde{A}=\tilde{D}^{-\frac{1}{2}}(A+I)\tilde{D}^{-\frac{1}{2}}$, $\tilde{D}$ represents the diagonal degree matrix after adding self-loop, $I$ indicates the identity matrix and $W$ denotes a trainable weight matrix. Here, $\sigma(\cdot)$ is the activation function.

To obtain remarkable performance, GNN optimizes the parameters $\theta$ by minimizing a cross-entropy loss function on the output of the training set $\mathcal{V}_{train}$. The corresponding cross-entropy loss function is provided as:
\begin{equation}
L(\theta;A,X)=-\sum_{v\in \mathcal{V}_{train}}\ln z_{v,c_v},
\end{equation}
where $\theta = \{W_0, \dots, W_k\}$ denotes the set of trainable weight matrices, $c_v$ denotes the ground truth label of node $v$, and $z_{v,c}$ is the probability of node $v$ belonging to class $c$.

\subsection{Graph Injection Attack}
For GIA, the original input is replaced with a perturbed graph $G^{\prime}=(A^{\prime}, X^{\prime})$ to degenerate the performance of GNN. Then, the optimization function of GIA is provided as:
\begin{equation}
\begin{split}
\min \sum_{v\in \mathcal{V}_{test}}\mathbb{I}(\arg \max \ln f_{\theta ^{*}}(A^{\prime}, X^{\prime})_{v}=c_v),\\
s.t.\ \theta^{*}=\arg\max\limits_{\theta}\sum_{v\in \mathcal{V}_{train}}\ln f_{\theta}(A, X)_{v,c_v},
\end{split}
\label{eq_giamin}
\end{equation}
where $\theta ^{*}$ are the optimal parameters of the GNN to be attacked and $\mathbb{I}(\cdot)$ is the indicator function. To minimize Eq. (\ref{eq_giamin}), the adversarial adjacency matrix $A^{\prime}$ is given as:
\begin{equation}
A^{\prime}= \begin{bmatrix}
A & B\\
B^{T} & Q\\
\end{bmatrix},   
\end{equation}
where $A$ indicates the original adjacency matrix. Instead of modifying $A$, GIA focuses on modifying $B\in \{0,1\}^{n\times n_{fake}}$ which indicates the connections among fake nodes and the nodes in $\mathcal{V}$, $n_{fake}$ indicates the number of fake nodes. Particularly, we assume $Q=O$ which indicates that there exists no link in any fake node pairs. As to the adversarial feature matrix $X^{\prime}$, the corresponding formula is given as:
\begin{equation}
X^{\prime}= \begin{bmatrix}
X\\
X_{fake}
\end{bmatrix},   
\end{equation}
where $X$ denotes the original feature matrix. Instead of modifying $X$, GIA focuses on modifying $X_{fake}\in \mathbb{R}^{n_{fake}\times m}$ which represents the feature matrix of fake nodes. For ease of future representation, Table \ref{tab_notation} provides some utilized notations and their interpretations.

\begin{table}[ht]
\caption{The descriptions of notations}
\centering
\begin{tabular}{ll}
\toprule
Notations & Descriptions \\
\midrule
$G$ & Original input graph \\
$\mathcal{V}$ & Node set of $G$ \\
$\mathcal{E}$ & Edge set of $G$ \\
$A$ & Adjacency matrix of $G$ \\
$X$ & Feature matrix of nodes in $G$ \\
$D$ & Degree matrix of $G$ \\
$d_i$ & Degree for node $i$ \\
$G^{\prime}=(A^{\prime}, X^{\prime})$ & Perturbed graph \\
$c$ & Set of the ground truth labels for nodes in $G$ \\
$c_b$ & Set of target labels for nodes in $G$ \\
$f_\theta$ & Graph neural network \\
$M_s$ & Surrogate model \\
$W$ & Weight matrix  \\
$n_{fake}$ & Number of fake nodes to be injected \\
$y_i$ & Predicted label for node $i$ \\
$Z$ & Predicted probability matrix \\
$h_i$ & Target label similarity for node $i$ \\
$s_h$ & Propagation scores for nodes in $G$ \\
$s_p$ & Cluster scores for potential clusters \\
$s_x$ & Feature scores for feature elements \\

\bottomrule
\end{tabular}
\label{tab_notation}
\end{table}

\section{Proposed Method}\label{sec_method}

In this section, we first discuss our motivation and observation from the perspective of label propagation. Based on those analyses, we derive the objective of our proposed method. Then, the details of our method are followed.

\subsection{Rethinking GIA from the perspective of label propagation}
Existing studies suggest that GNN and label propagation are related with a learnable feature mapping \cite{appna2019,pta2021,gcnlpa2021}. Hence, GNN can be seen as a scheme that obtains the initial prediction from the mapping and then propagates the prediction in the graph using label propagation algorithms \cite{cs2021}. Following this scheme, we attempt to estimate the probability distribution changes for nodes before and after the GIA using the label propagation algorithm. Specifically, inspired by \cite{llg2003,cs2021}, the adopted label propagation process for analyzing GIA is expressed as:
\begin{equation}
Z^{t+1} = \alpha\cdot D^{-\frac{1}{2}}AD^{-\frac{1}{2}}Z^{t} + (1-\alpha)Z^{0},
\label{eq_z}
\end{equation}
where $Z\in \mathbb{R}^{n\times c}$ denotes a matrix in which each row represents a probability distribution resulting from the softmax and $\alpha$ denotes the hyper-parameter. 

Then, to conduct GIA, we suppose that fake node $u$ with a one-hot probability distribution is connected to node $i$ in the original graph aiming to change the probability distribution of node $i$. Thus, for the victim node $i$, its probability distribution after the attack is given as:
\begin{equation}
z^{t+1}_{i} = \alpha\cdot \sum_{j\in \mathcal{V}_i}\frac{1}{\sqrt{d_i+1}\sqrt{d_j}}z^{t}_j + \alpha\cdot \frac{1}{\sqrt{d_i+1}\sqrt{d_u}}z^{t}_u + (1-\alpha)z^{0}_i,
\label{eq_z_t+1_i}
\end{equation}
where $\mathcal{V}_i$ is the original neighbor set of $i$ and $d_i$ indicates the original degree of the victim node $i$. In particular, according to Eq. (\ref{eq_z_t+1_i}), we reveal that the above injection operation increases the probability of predicting the victim nodes as the injected node's label rather than other categories. Thus, to misclassify node $i$, the objective of injection can be expressed as:
\begin{equation}
\max(z^{t+1}_{i,c_u}-z^{t+1}_{i,y_i}),
\label{eq_maxzi}
\end{equation}
where $y_i=\arg\max z^{0}_{i}$ denotes the initially predicted label of node $i$. If $z^{t}_{i,c_u}-z^{t}_{i,y_i}>0$, we suppose that node $i$ is classified as the injected node's label $c_u$. For the ease of altering the probability ranking of categories, we choose $c_{u}=\arg\max_{c\neq y_i} z^0_{i,c}$ as the optimal pseudo label for the injected node $u$ which is the second largest probability category for node $i$. Along this line, we also define $c_{b,i}=\arg\max_{c\neq y_i} z^0_{i,c}$ as the target label of the victim node $i$. Since the target label has been determined for each node in the original graph, we can transform the global GIA problem into a global injection label specificity attack problem.

{\bf{Definition 1}} (Global Injection Label Specificity Attack Problem). Given a graph $G=(A, X)$, the global injection label specificity attack problem is to find a perturbed graph $G^{\prime}=(A^{\prime}, X^{\prime})$ through injecting fake nodes so that nodes in the original graph could be classified as their target labels after applying GNN $f_\theta$ on $G^{\prime}$. Therefore, the objective in Eq. (\ref{eq_giamin}) can be reinterpreted as:
\begin{equation}
\begin{split}
\max \sum_{v\in \mathcal{V}_{test}}\mathbb{I}(\arg \max \ln f_{\theta ^{*}}(A^{\prime}, X^{\prime})_{v}=c_{b,v}),\\
s.t.\ \theta ^{*}=\arg\max\limits_{\theta}\sum_{v\in \mathcal{V}_{train}}\ln f_{\theta}(A, X)_{v,c_v},
\end{split}
\label{eq_giamaxcb}
\end{equation}
where $c_{b,v}$ is the target label for a victim node $v$.

To solve this problem, our attack needs to increase the probability of the victim nodes belonging to the corresponding target label through propagating the probability distribution. When the probability distribution further propagates from the victim node to its neighbors, a crucial question is whether there exists a similarity among target labels of the  neighboring nodes. The presence of target label similarity between neighboring nodes facilitates the reuse of malicious features to attack the neighboring nodes of the victim node through propagation. On the contrary, the dissimilarity will turn our global attack into an inefficient single-node attack. Inspired by \cite{geom2020,h2gcn2020,gani2022}, for node $i$, the target label similarity can be calculated as:
\begin{equation} 
h_i = \frac{\sum_{j\in \mathcal{V}_i}\mathbb{I}(c_{b,j}=c_{b,i})}{d_i},
\label{eq_hi}
\end{equation}
where $\mathcal{V}_i$ is the neighbor set of $i$ and $d_i$ denotes the degree of node $i$. As shown in Table \ref{tab_tarhomo}, the target labels derived from various GNNs demonstrate a similarity similar to the homophily assumption. This property indicates that our attack on one node can also help achieve attacks on its neighboring nodes, enhancing the feasibility of problem transformation.

\begin{table}[htbp]
\caption{Similarity based on the target label in the homophily graph.}
\centering
\begin{tabular}{ccccc}
\toprule
Similarity & Cora   & Cora\_ML & Citeseer & Pubmed \\
\midrule
GCN Target Label              & 0.69 & 0.66   & 0.76   & 0.81 \\
SGC Target Label              & 0.64 & 0.60   & 0.71   & 0.75 \\
GAT Target Label              & 0.69 & 0.69   & 0.77   & 0.84 \\
\bottomrule
\end{tabular}
\label{tab_tarhomo}
\end{table}

\begin{figure*}[htbp]
    \centering
    \includegraphics[width=7in]{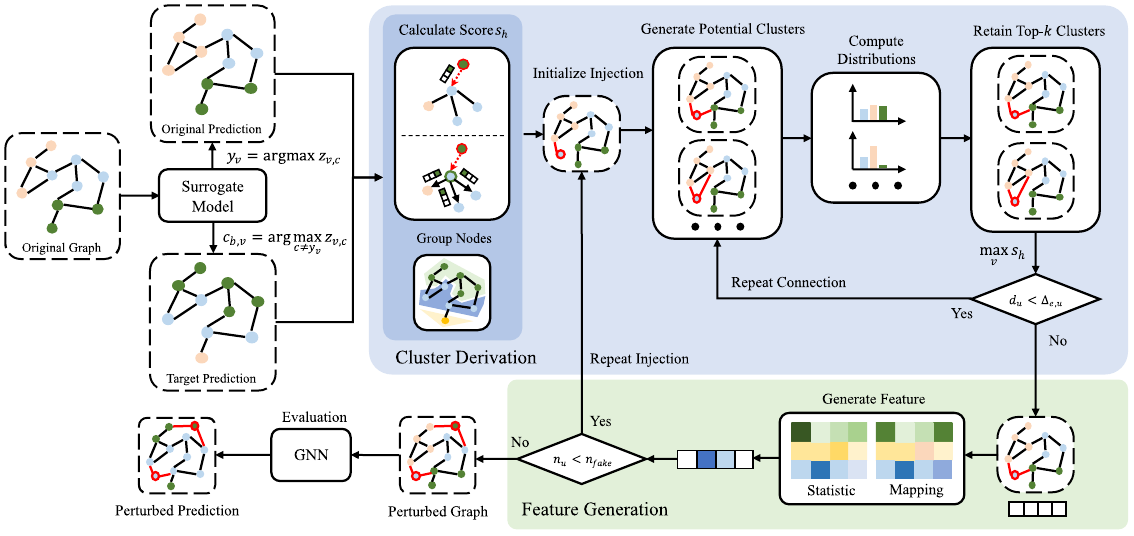}
    \caption{The illustration of LPGIA. We obtain the target label for each node and group the nodes accordingly. Then, for an injected node, the node with the largest score $s_h$ is selected as the initial node to be attacked. Iteratively, the top-$k$ nodes are retained according to the target label's probability in the aggregated smooth distribution, and then the one with the largest score $s_h$ is selected. Finally, we leverage the feature mapping and the statistics to generate the malicious feature for the injected node.}
    \label{fig_intro}
\end{figure*}

To address the global injection label specificity attack problem, we propose the label-propagation-based global injection attack (LPGIA), leveraging the similarity of the target label between neighboring nodes. Our proposed method is composed of two parts, i.e., cluster derivation and feature generation. In the cluster derivation procedure, for each fake node, LPGIA heuristically determines the optimal victim cluster connected to it. In the feature generation procedure, LPGIA generates malicious features in order to increase the probability of the target label. An illustration of the proposed LPGIA is shown in Fig. \ref{fig_intro}.

\subsection{Cluster Derivation}
First and foremost, we obtain the target label for each node in the original graph. Specifically, a GNN-based surrogate model $M_s$ is trained to obtain the target label from the prediction $Z=M_s(G)$. Inspired by \cite{cs2021}, to reduce the uncertainty of the prediction from GNN, Eq. (\ref{eq_z}) is applied to the original prediction $Z$ to get the smooth probability distribution $\tilde{Z}$ until convergence. Thus, we derive the stable target label $c_{b,v} = \arg\max_{c\neq y_v} \tilde{z}_{v,c}$ for the node $v$. Later, we classify all the potential victim nodes into a number of groups according to the target label. The next step is to find an optimal combination of victim nodes in the group for the injected nodes. Specifically, we propose a heuristic strategy that involves the identification of vulnerable and critical nodes in the graph, followed by clustering these nodes.

{\bf{Identifying Valuable Nodes.}}
During the node identification stage, the neighborhood and the feature of the injected node are not yet determined. Thus, it is hard to estimate the probability distribution propagating to the victim node from the injected node. However, we assume that the pseudo-label of the injected node is the target label of the connected victim node. Based on this situation, we use the idea of hard-label propagation to identify valuable nodes to be attacked. In hard-label propagation, the label of a node depends on the label occurring with the highest frequency among its neighboring nodes \cite{lpch2007}. This propagation process is depicted as:
\begin{equation}
y^t_i = \arg\max\limits_{c} (\frac{\sum_{j\in \mathcal{V}_i}\mathbb{I}(y^t_j=l_1)}{d_i}, \dots, \frac{\sum_{j\in \mathcal{V}_i}\mathbb{I}(y^t_j=l_L)}{d_i}),
\label{eq_hlp}
\end{equation}
where $l_1$ is the first category in the label set $l$, $y_j$ is the predicted label of node $j$ and $t$ represents the number of iterations for the propagation process.

For a victim node $i$, injecting a node can change the frequency of its original label among its neighboring nodes, i.e., $\frac{\sum_{j\in \mathcal{V}_i}\mathbb{I}(y_j=y_i)}{d_i+1}$. Intuitively, a node that exhibits a significant decrease in the proportion of its original label among its neighbors' labels when subjected to an attack is relatively more susceptible to perturbation. Therefore, the vulnerability weight of the node $i$ can be defined as:
\begin{equation}
s_{1_i} = \frac{\sum_{j\in \mathcal{V}_i}\mathbb{I}(y_j=y_i)}{d_i^2+d_i},
\label{eq_s1}
\end{equation}
where $\mathcal{V}_i$ is the neighbor set of node $i$. $s_{1_i}$ measures the decreasing extent of the proportion of its initially predicted label among its neighbors' labels under an attack. A larger value of $s_{1_i}$ indicates that node $i$ is more susceptible to adversarial perturbation.

Then, suppose we successfully classify node $i$ with its target label $c_{b, i}$. In that case, its neighboring nodes will tend to be misled to the same label, i.e., $c_{b, i}$. To utilize the propagation process to achieve our goal in Eq. (\ref{eq_giamaxcb}), we shall attack the node $i$ with a higher similarity of the target label between neighbors, i.e., $\frac{\sum_{j\in \mathcal{V}_i}\mathbb{I}(c_{b, j}=c_{b, i})}{d_i}$. Therefore, the topological weight of the node $i$ can be defined as:
\begin{equation}
s_{2_i} = \frac{\sum_{j\in \mathcal{V}_i}\mathbb{I}(c_{b,j}=c_{b,i})-\sum_{j\in \mathcal{V}_i}\mathbb{I}(y_j=c_{b,i})}{d_i},
\label{eq_s2}
\end{equation}
where $c_{b,i}$ indicates the target label of the victim node $i$.

Based on the above analyses, for the victim node $i$, the propagation score is defined to comprehensively evaluate the impact of the node injection, which combines $s_{1_i}$ and $s_{2_i}$. The formula is provided as:
\begin{equation}
s_{h_i}=\beta \cdot s_{1_i}+(1-\beta)\cdot s_{2_i},
\label{eq_s_h}
\end{equation}
where $\beta$ is the hyper-parameter.

\begin{figure}[htbp]
    \centering
    \includegraphics[width=3in]{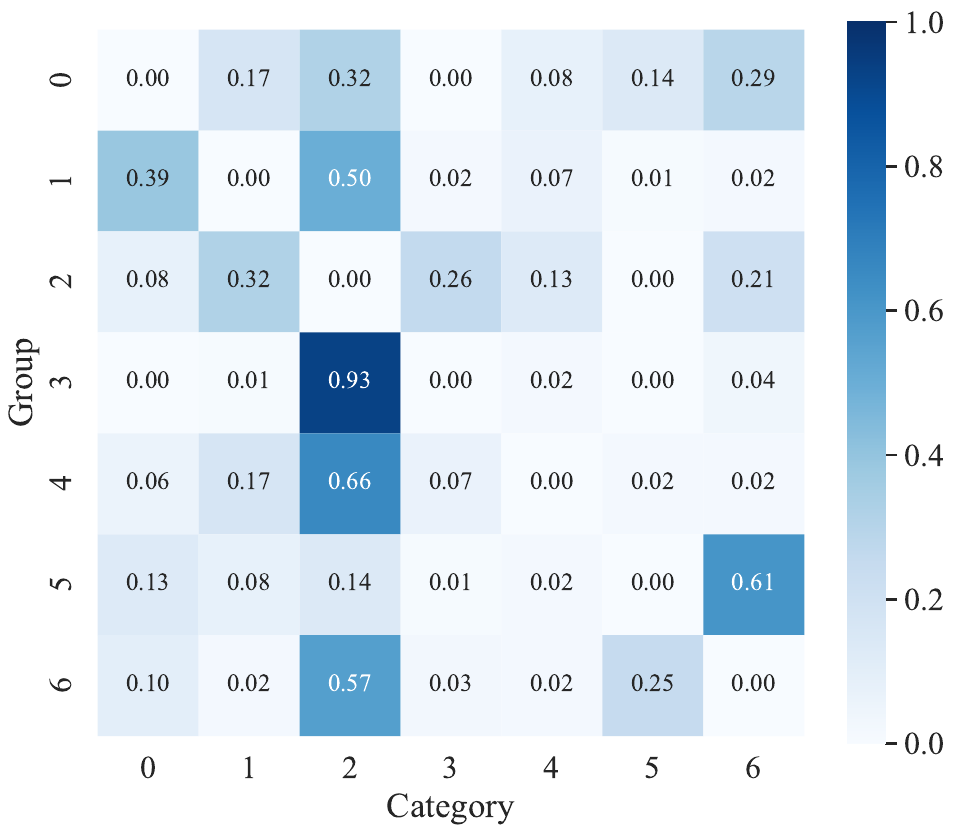}
    \caption{Frequency of categories in each target label group. The example is derived from the prediction of GCN on the Cora dataset. This phenomenon also exists in the predictions of other GNNs applied to homophily graphs.}
    \label{fig_imbalance}
\end{figure}

{\bf{Clustering Nodes.}}
According to the propagation score, we can propose a simple clustering strategy. We rank the nodes in each target label group respectively and select the top-$k$ nodes from one of the groups to form a cluster, where $k$ is the connection budget for a given injected node $u$. Then, we assign the target label to the injected node $u$ as its pseudo label $c_u$ and connect it to the nodes in the cluster. However, this strategy has a drawback. There may exist a category imbalance phenomenon, where the majority of the nodes in a target label group belong to the same original label, as shown in Fig. \ref{fig_imbalance}. Thus, there is a high probability of selecting several nodes with the same original label into a cluster. In this case, the injected node will aggregate probability distribution having a high probability of the non-target label according to Eq. (\ref{eq_z}), and then propagate it back to its neighbors. Obviously, it is contrary to our objective mentioned in Eq. (\ref{eq_maxzi}).

To alleviate this issue, we attempt to cluster nodes in a way that maximizes the probability of the target label in the aggregated probability distribution of the corresponding injected node while considering the propagation score of the node. Then, according to Eq. (\ref{eq_z}), the aggregated probability distribution can be calculated as:
\begin{equation}
z^{t+1}_{u} = \sum_{i\in \mathcal{V}_u}\frac{1}{\sqrt{d_i+1}\sqrt{d_u}}z^{t}_i,
\label{eq_z_t+1_u}
\end{equation}
where $\mathcal{V}_u$ is the neighbor set of injected node $u$. Later, the objective is expressed as $\max z^{t+1}_{u,c_u}$. To avoid excessive computation, we simplify the above propagation process and obtain an approximate result through performing propagation once. In particular, we propagate the smooth probability distribution $\tilde{Z}$ instead of the original distribution to reduce the loss of information. Thus, the objective can be reformulated as $\max \tilde{z}^{1}_{u,c_u}$.


To achieve the above goal, we construct the cluster $p$ through a greedy algorithm. Specifically, the node with the largest propagation score is selected as the initial node of the victim cluster $p$, and then nodes with the same target label become potential options for the following selections. Iteratively, we add a new victim node into the cluster $p$. As our goal is to maximize the target label's probability in the aggregated smooth distribution, we define a cluster score $s_p$ to measure the difference between the probability of the target label and the maximum probability among other labels after adding each potential node to the cluster. First, for node $j$, according to Eq. (\ref{eq_z_t+1_u}), the aggregated smooth probability distribution after adding is calculated as: 
\begin{equation}
\tilde{z}^{\prime}_{p_j}=\frac{1}{\sqrt{d_u}}(\sum_{i\in \mathcal{V}_p}\frac{1}{\sqrt{d_i+1}}\tilde{z}_{i}+\frac{1}{\sqrt{d_j+1}}\tilde{z}_{j}),
\label{eq_z_j}
\end{equation}
where $\mathcal{V}_p$ is composed of nodes that are selected to form the cluster, $\tilde{z}_{i}$ indicates the smooth probability distribution of node $i$ and $u$ represents the injected node being assigned to this cluster $p$. Then, the corresponding cluster score is defined as:
\begin{equation}
\begin{split}
s_{p_j} = \tilde{z}^{\prime}_{p_j,c_{u}}-\tilde{z}^{\prime}_{p_j,c^*}, \\
s.t.\ c^{*}=\arg\max\limits_{c\neq c_{u}}\tilde{z}^{\prime}_{p_j},
\end{split}
\label{eq_s_p}
\end{equation}
where $\tilde{z}^{\prime}_{p_j,c_{u}}$ is the probability of the target label in the distribution. After computing the cluster scores, the potential nodes can be ranked accordingly; then, we retain the top-$n_k$ nodes, which alleviates the aforementioned imbalance problem. Nevertheless, a large cluster score does not necessarily mean that corresponding node is valuable to attack. To better leverage the similarity of the target label between neighboring nodes, we finally add the node with the largest propagation score $s_h$ among the retained nodes into the victim cluster. 

Finally, we remove the selected node from the corresponding group aiming to ensure that each selected node will only be connected to one injected node during the attack. The above process is repeated until the link budget of a single injected node $\Delta_{e,u}$ has been exhausted. Similar to \cite{afgsm2020,gani2022}, we adopt a degree sampling operation to assign the link budget $\Delta_{e,u}$ to each fake node $u$. They argue that the sampling operation can reduce the damage to the degree distribution of the original graph during the attack process, leading to a more unnoticeable attack. Furthermore, we let the total number of the sampling degree $\sum \Delta_{e,u} = \lfloor n_{fake} \times \frac{\sum_{v\in\mathcal{V}}d_v}{n} \rfloor$. This step is conducted aiming to reduce the randomness and achieve fair comparison. 

\subsection{Feature Generation}
In the feature generation procedure, we are anticipated to generate the malicious feature for each injected node. To achieve our goal, for an injected node $u$, the key point of designing the malicious feature is to increase the probability of the target label in its probability distribution. Similar to \cite{gnia2021,cs2021}, we can estimate the probability distribution before propagation using the feature mapping. Thus, the objective of designing the malicious feature can be expressed as:
\begin{equation}
\max \sum_{k=0}^{m} x_{u,k}W_{k,c_u}, 
\label{eq_tlpmax}
\end{equation}
where $W_{k,c_u}$ is the weight of the $k$-th feature element for target label $c_u$ and $m$ is the dimension of the node feature $x_u$.

Notably, it is unrealistic to arbitrarily set the values of feature elements. Inspired by \cite{gani2022}, we apply a statistical method to meet the unnoticeable requirement. Hence, for an injected node $u$, the possible value of the feature element is provided as: 
\begin{equation}
x^{\prime}_{u,k} = \frac{\sum_{i\in \mathcal{V}_{c_u}}|x_{i,k}|}{\sum_{i\in \mathcal{V}_{c_u}}\mathbb{I}(x_{i,k}\neq 0)},
\label{eq_x_c}
\end{equation}
where $\mathcal{V}_{c_u}$ denotes the set of nodes possessing the label same as the injected node label $c_u$ and $x_{i,k}$ denotes the $k$-th element of node feature $x_i$. Accordingly, we can generate features that are in the same range as the original ones. Furthermore, this operation is applicable to graphs with binary or continuous feature spaces.

Then, to obtain the weight of feature elements, following previous works \cite{afgsm2020,gnia2021,sga2021}, we perform a linearization of the surrogate model, i.e., $W=W_0\dots W_{k}$, where $k$ is the number of the weight matrix. To alleviate the overfitting problem in the mapping, we train the surrogate model for several times and then derive the average of those linearization weights which is regarded as the final weight matrix $\overline{W}$.

Once we have obtained the possible value and weight of the feature elements, we can determine their values to achieve the goal in Eq. (\ref{eq_tlpmax}). Additionally, we also restrict the probability of other labels. Hence, for a feature element $x_{u,k}$ of the injected node $u$, the corresponding feature element score is computed as:
\begin{equation}
\begin{gathered}
s_{x_k} = (\overline{W}_{k,c_u}-\overline{W}_{k,c_z})\cdot x^{\prime}_{u,k}, \\
s.t.\ c_z=\arg\max\limits_{c\neq c_{u}}\sum_{i\in \mathcal{V}_u}\frac{1}{\sqrt{d_i+1}}\tilde{z}_{i},
\end{gathered}
\label{eq_s_x}
\end{equation}
where $\overline{W}_{k,c_u}$ is the average weight of the $k$-th feature element for target label $c_u$ and the set $\mathcal{V}_u$ is composed of connected victim nodes. Simply, a feature element $x_{u,k}$ can be selected if $s_{x_k}>0$. Then, the values of the unselected feature elements are set to 0. Nevertheless, the feature space might be sparse in some graphs. Similar to \cite{afgsm2020,gnia2021,gani2022}, there is a constraint on the number of non-zero elements of each malicious feature. Then, the budget for a new feature vector is given as:
\begin{equation}
\Delta_x=\lfloor \frac{\sum_{v\in \mathcal{V}}\sum^{m}_{k=0}\mathbb{I}(x_{v,k}\neq 0)}{n} \rfloor.
\label{eq_delta_x}
\end{equation}

Aiming to increase the diversity of malicious features, we perform a random selection operation instead of directly selecting the top-$\Delta_x$ feature elements according to the score being calculated in Eq. (\ref{eq_s_x}). Specifically, considering the limitation of allocated budgets, we randomly select $\Delta_x$ feature elements from the top-$\min(2\Delta_x, m)$ elements to generate the malicious feature. Eventually, the malicious feature of the injected node $u$ can be depicted as:
\begin{equation}
x_{u,k}=\left\{
\begin{aligned}
&x^{\prime}_{u,k} &,& \text{ if } x_{u,k} \text{ is selected}, \\
&0 &,& \text{ if } x_{u,k} \text{ is not selected}.
\end{aligned}
\right.
\label{eq_x_u}
\end{equation}

\subsection{Overall Attack Process}
Overall, the framework of our proposed LPGIA is summarized in Algorithm 1. Based on our analysis, we transform the GIA problem into a global injection label specificity attack problem. Then, we conduct sequential injection for the graph. For each fake node, we select a victim node to determine the pseudo label of the fake node and the potential nodes for selection. Iteratively, we rank the nodes according to the cluster score; and then the one with the largest propagation score in the top-$n_k$ nodes is connected to the fake node. After the budget for a fake node is exhausted, we leverage the feature mapping and the statistic to generate malicious features for the fake node. The above processes are repeated until all the fake nodes are injected.

\begin{algorithm}
\caption{Label-propagation-based Global Injection Attack}
\begin{algorithmic}[1]
\REQUIRE Original graph $G=(A,X)$; surrogate model $M_s$; the victim node set $\mathcal{V}_a$; the number of fake nodes $n_{fake}$; hyper-parameters $\beta$ for calculating propagation score; the coefficient of label propagation $\alpha$.
\ENSURE Adversarial graph $G^\prime = (A^\prime,X^\prime)$
\STATE Get average linearized weight $\overline{W}$ and original prediction $Z^{0}$ using $M_s(G)$.
\STATE $\Delta_e \leftarrow $ Generate the link budget of each fake node based on degree sampling operation.
\STATE $\Delta_x \leftarrow $ $\lfloor \frac{\sum_{v\in \mathcal{V}}\sum^{m}_{k=0}\mathbb{I}(x_{v,k}\neq 0)}{n} \rfloor$.
\STATE $x^{\prime}_{c,k} \leftarrow $ Obtain the possible value of feature elements according to Eq. (\ref{eq_x_c}).
\STATE $\tilde{Z} \leftarrow $ Iterate $Z^{t+1} = \alpha\cdot D^{-\frac{1}{2}}AD^{-\frac{1}{2}}Z^{t} + (1-\alpha)Z^{0}$ until convergence to $\tilde{Z}$.
\FOR{$i=0$ to $n$}
\STATE $c_{b,i} \leftarrow $ $\arg\max_{c\neq y_i} z_{i}^{0}$.
\ENDFOR
\STATE Group victim nodes by the target label $c_b$.
\STATE $s_h \leftarrow $ Compute propagation scores for each node in $\mathcal{V}_a$ according to Eq. (\ref{eq_s_h}).
\FOR{$u=0$ to $n_{fake}$}
\STATE Initialize the cluster $p_u$ via the ranking of the propagation score $s_h$.
\FOR{$j=1$ to $\Delta_{e,u}$}
\STATE Calculate the cluster score $s_p$ for all potential nodes according to Eq. (\ref{eq_s_p}).
\STATE Rank the nodes according to score $s_p$ and retain the top-$n_k$ nodes.
\STATE Add the node with the largest propagation score $s_h$ in the retained nodes into the cluster and remove it from potential nodes.
\ENDFOR
\STATE $x_{u} \leftarrow $ Generate the adversarial feature vector according to Eq. (\ref{eq_x_u}).
\STATE Update $A^\prime,X^\prime$ according to $p_{u}, x_{u}$.
\STATE $Z^\prime \leftarrow $ $M_s(G^\prime)$
\STATE $\tilde{Z^\prime} \leftarrow $ Iterate $Z^{t+1} = \alpha\cdot D^{-\frac{1}{2}}A^{\prime}D^{-\frac{1}{2}}Z^{t} + (1-\alpha)Z^{\prime}$ until convergence to $\tilde{Z^\prime}$.
\ENDFOR
\RETURN Adversarial graph $G^\prime = (A^\prime,X^\prime)$
\end{algorithmic}
\label{alg1}
\end{algorithm}

\section{Experiments}\label{sec_exp}
In this section, the superiority of our proposed LPGIA is verified through comparing the corresponding performance with that of four GIA baselines. The experiments are conducted on five representative datasets against five GNNs on the node classification task. Furthermore, ablation studies are also performed for discussions. Our code is available at https://github.com/NPU-Netsci/LPGIA.

\subsection{Experimental Settings}
The considered five representative datasets are listed as Cora, Citeseer, Pubmed \cite{planetoid2016}, Cora-ML \cite{coraml2018}, and OGB-Arxiv \cite{ogb2020}. For OGB-Arxiv, we use the official data loader to split it. For other datasets, we randomly split the node set into three parts, including the training set (10\%), validation set (10\%), and testing set (80\%). Accordingly, we applied some limitations for node injection. For each dataset, we set up the budget on the number of total new connections which equals to the product of the average node degree and the number of fake nodes. Besides, we also limit the number of non-zero feature elements of the fake nodes to avoid generating abnormal features. For Cora\_ML, Pubmed, and OGB-Arxiv, their non-zero feature elements obey a skewed distribution and we use the 99th percentile as a limitation of the value range. Following \cite{gnia2021,gani2022}, we only retain the largest connected component for subsequent experiments, with the corresponding statistics and injection constraints being presented in Table \ref{tab_dataset}.

\begin{table}[t]
\caption{Statistics of the considered datasets. Here, ANF represents the average number of non-zero feature elements in a feature vector and IFR represents the feature element range of the injected nodes.}
\centering
\scalebox{0.85}{
\begin{tabular}{cccccccc}
\toprule
\makebox[0.01\textwidth][c]{Dataset} & \makebox[0.01\textwidth][c]{Nodes} & \makebox[0.01\textwidth][c]{Edges} & \makebox[0.01\textwidth][c]{Labels} & \makebox[0.04\textwidth][c]{Features}& \makebox[0.01\textwidth][c]{ANF}& \makebox[0.01\textwidth][c]{IFR}& \makebox[0.05\textwidth][c]{Avg. Degree}\\
\midrule
Cora     & 2485  & 5069  & 7      & 1433     & 18.3       & \{0, 1\}      & 4.1        \\
Cora\_ML & 2810  & 7981  & 7      & 2879     & 50.6       & [0, 0.42]     & 5.7        \\
Citeseer & 2110  & 3668  & 6      & 3703     & 32.1       & \{0, 1\}      & 3.5        \\
Pubmed   & 19717 & 44324 & 3      & 500      & 50.1       & [0, 0.22]     & 4.5        \\
OGB-Arxiv & 169343 & 1157799 & 40 & 128      & 128.0       & [-0.48, 0.80] & 13.7        \\

\bottomrule
\end{tabular}}
\label{tab_dataset}
\end{table}

In order to evaluate the attacking performance and transferability of our method, we consider five well-known GNNs to be attacked, which represent different types of neighborhood aggregation, including GCN \cite{gcn2017}, GAT \cite{gat2018}, SGC \cite{sgc2019}, GNNGuard \cite{gnnguard2020} and SimPGCN \cite{simpgcn2021}. To illustrate the superiority, the performance of our LPGIA is compared with those of four GIA methods including AFGSM, G-NIA, ClusterAttack and GANI. Among these methods, AFGSM and G-NIA are two targeted attack methods. To extend them to conduct global attack, we randomly select a target node, then optimize the edge and feature using the original method respectively. The above process is repeated until all fake nodes are injected. Details of the considered baselines are provided:
\begin{enumerate}[(1)]
\item AFGSM\cite{afgsm2020}: AFGSM is a gradient-based targeted attack method. It perturbs the edges or features according to the approximate closed-form solution. Nevertheless, the operation only satisfies the binary feature space. For continuous feature space, we replace 1 with the maximum value of the range mentioned in Table \ref{tab_dataset}.

\item G-NIA\cite{gnia2021}: Similar to AFGSM, G-NIA is also a gradient-based method focusing on the targeted attack. It trains multilayer perceptrons in order to learn the attack strategy. As the multilayer perceptron generates continuous output, G-NIA extends the Gumbel-Softmax approach to obtain the discrete features and edges.

\item ClusterAttack\cite{ca2022}: ClusterAttack transforms the GIA problem into a graph clustering problem. It clusters the victim nodes according to Euclid's distance between their adversarial feature vectors. Subsequently, it uses zeroth-order optimization to optimize the injected node feature for the graph with binary features. For continuous features, it utilizes natural evolution strategies instead.

\item GANI\cite{gani2022}: GANI introduces the genetic algorithm to optimize potential edge combinations and uses the total decrease of node homophily to sort the combinations further. Moreover, it employs statistical operations to generate malicious features.
\end{enumerate}

In this manuscript, the experiments are conducted for 50 times and the average results will be reported to reduce the randomness. The experiments follow the black-box settings in previous works \cite{afgsm2020,gnia2021,gani2022}, where the attacker has no information about the defense model. Thus, the attacker needs to utilize a common GNN as a surrogate model. For our proposed method, we train a GCN with 2 layers as our surrogate model and this is performed for 20 times to derive the average of linearization weights. In the label propagation-based smooth operation, coefficient $\alpha$ is 0.9, and the maximum number of iterations $T$ is 50. The coefficient of propagation score $\beta$ in Eq. (\ref{eq_s_h}) is set as 0.5 by default. Furthermore, the number of retained potential nodes $n_k$ equals to 10. 

\subsection{Performance of Attack Methods on Small Datasets}
Similar to \cite{hao2022,gani2022}, we evaluate the attacking performance of LPGIA by comparing it with other baselines under a 5\% injected ratio. According to the attack occurring stages, adversarial attacks can be categorized into two types, i.e., poisoning attack and evasion attack, for which the attack occurs during the training and testing stages of GNN respectively. Thus, we investigate the corresponding performance of our LPGIA under the different attacks. 

\begin{table}[tbp]
\caption{Accuracy (\%) of the GNNs under evasion attack on small datasets. Here, a smaller value usually indicates greater damage incurred by the attack. Avg. is the average accuracy of five victim models on one dataset. The best results are boldfaced.}
\centering
\scalebox{0.8}{
\begin{tabular}{cccccccc}
\toprule
Dataset                   & Victim Model & Clean & AFGSM & G-NIA & CA    & GANI  & LPGIA            \\ \midrule
\multirow{6}{*}{Cora}     & GCN          & 83.93 & 79.94 & 81.77 & 81.34 & 77.77 & \textbf{75.35} \\
                          & SGC          & 83.83 & 78.49 & 81.26 & 81.79 & 78.18 & \textbf{75.16} \\
                          & GAT          & 83.30 & 80.38 & 81.53 & 82.10 & 78.27 & \textbf{75.80} \\
                          & GNNGuard     & 83.65 & 79.91 & 81.52 & 81.99 & 78.57 & \textbf{74.55} \\
                          & SimPGCN      & 82.35 & 79.92 & 80.97 & 80.23 & 78.83 & \textbf{74.44} \\
                          & Avg.          & 83.41 & 79.73 & 81.41 & 81.49 & 78.32 & \textbf{75.06} \\ \midrule
\multirow{6}{*}{Cora\_ML} & GCN          & 85.99 & 83.23 & 83.85 & 84.16 & 79.80 & \textbf{73.11} \\
                          & SGC          & 84.09 & 82.74 & 83.54 & 83.63 & 80.96 & \textbf{73.09} \\
                          & GAT          & 85.39 & 83.25 & 83.70 & 84.34 & 81.14 & \textbf{73.28} \\
                          & GNNGuard     & 85.76 & 83.96 & 83.51 & 83.85 & 81.03 & \textbf{73.39} \\
                          & SimPGCN      & 86.19 & 84.63 & 83.44 & 84.21 & 81.20 & \textbf{73.46} \\
                          & Avg.          & 85.48 & 83.56 & 83.61 & 84.04 & 80.83 & \textbf{73.27} \\ \midrule
\multirow{6}{*}{Citeseer} & GCN          & 72.63 & 69.98 & 69.73 & 70.63 & 69.69 & \textbf{66.23} \\
                          & SGC          & 72.49 & 70.67 & 70.50 & 70.77 & 69.61 & \textbf{66.00} \\
                          & GAT          & 72.46 & 71.72 & 70.32 & 70.76 & 71.19 & \textbf{67.66} \\
                          & GNNGuard     & 72.75 & 71.54 & 71.37 & 70.74 & 70.56 & \textbf{66.17} \\
                          & SimPGCN      & 74.50 & 71.92 & 71.56 & 70.91 & 70.54 & \textbf{69.12} \\
                          & Avg.          & 72.96 & 71.17 & 70.70 & 70.76 & 70.32 & \textbf{67.04} \\ \midrule
\multirow{6}{*}{Pubmed}   & GCN          & 86.26 & 84.92 & 85.52 & 82.65 & 82.57 & \textbf{75.43} \\
                          & SGC          & 82.21 & 80.09 & 79.99 & 79.36 & 80.41 & \textbf{73.82} \\
                          & GAT          & 85.57 & 84.13 & 84.77 & 81.70 & 82.16 & \textbf{74.61} \\
                          & GNNGuard     & 86.73 & 85.48 & 86.05 & 83.30 & 82.81 & \textbf{75.74} \\
                          & SimPGCN      & 87.91 & 87.11 & 86.72 & 85.83 & 86.90 & \textbf{79.64} \\
                          & Avg.          & 85.73 & 84.35 & 84.61 & 82.57 & 82.97 & \textbf{75.85} \\
\bottomrule                          
\end{tabular}}
\label{tab_res_eva_5}
\end{table}

\subsubsection{Evasion Attack}
Table \ref{tab_res_eva_5} shows the classification accuracy of the victim GNNs on node classification tasks under evasion attack. Here, Clean indicates the result obtained from the GNN before attacks and CA is short for ClusterAttack. As revealed, for all the datasets, our LPGIA is of superior attack performance (being indicated by the smallest GNN Accuracy) compared with considered attack methods. As to the binary feature datasets, i.e., Cora and Citeseer, LPGIA can improve the performance by 3\% compared with the best baseline. Whereas for the continuous feature datasets, i.e., Cora\_ML and Pubmed, the improved amount is approximately 7\%. Thus, we can conclude that in the continuous feature space, LPGIA is of superior capability to boost the attack performance. This might be incurred by the fact that we consider the feature element values during feature generation.

Similar to other baselines \cite{afgsm2020,gnia2021,ca2022,gani2022}, LPGIA adopts GCN as the surrogate model and generates perturbations based on it. The perturbations will be used to attack other GNNs. Upon this setting, the results also indicate that the perturbations generated by LPGIA exhibit excellent transferability. Although GCN is adopted as the surrogate model, we observed that SGC performs worse compared with GCN when LPGIA is conducted. The absence of a nonlinear activation layer in SGC may decrease its robustness. SimPGCN exhibits the highest level of robustness among the tested GNN models in most cases, especially in Citeseer and Pubmed. As GIA methods avoid modifying the original graph, it is hard for them to perturb the process of constructing new graphs in SimPGCN. The new graphs could help SimPGCN learn more reliable node representations compared with other GNN models.
\begin{table}[tbp]
\caption{Accuracy (\%) of the GNNs under poisoning attack on small datasets.}
\centering
\scalebox{0.8}{
\begin{tabular}{cccccccc}
\toprule
Dataset                   & Victim Model & Clean & AFGSM & G-NIA & CA    & GANI  & LPGIA            \\ \midrule
\multirow{6}{*}{Cora}     & GCN          & 83.93 & 80.88 & 82.71 & 81.14 & 77.51 & \textbf{76.51} \\
                          & SGC          & 83.83 & 80.03 & 82.28 & 81.29 & 77.54 & \textbf{76.57} \\
                          & GAT          & 83.30 & 81.92 & 82.75 & 82.39 & 80.56 & \textbf{77.59} \\
                          & GNNGuard     & 83.65 & 80.72 & 82.63 & 81.64 & 78.31 & \textbf{76.41} \\
                          & SimPGCN      & 82.35 & 79.90 & 81.88 & 80.84 & 78.02 & \textbf{77.26} \\
                          & Avg.          & 83.41 & 80.69 & 82.45 & 81.46 & 78.39 & \textbf{76.87} \\ \midrule
\multirow{6}{*}{Cora\_ML} & GCN          & 85.99 & 82.92 & 84.56 & 84.14 & 80.20 & \textbf{76.42} \\
                          & SGC          & 84.09 & 82.90 & 83.59 & 82.78 & 80.65 & \textbf{76.42} \\
                          & GAT          & 85.39 & 83.32 & 85.01 & 84.83 & 82.21 & \textbf{78.16} \\
                          & GNNGuard     & 85.76 & 83.19 & 84.79 & 84.61 & 81.25 & \textbf{75.85} \\
                          & SimPGCN      & 86.19 & 83.74 & 84.83 & 84.30 & 80.74 & \textbf{75.18} \\
                          & Avg.          & 85.48 & 83.21 & 84.56 & 84.13 & 81.01 & \textbf{76.41} \\ \midrule
\multirow{6}{*}{Citeseer} & GCN          & 72.63 & 70.91 & 71.70 & 70.82 & 70.27 & \textbf{66.48} \\
                          & SGC          & 72.48 & 70.73 & 71.90 & 70.58 & 69.81 & \textbf{66.77} \\
                          & GAT          & 72.46 & 71.09 & 72.24 & 70.57 & 71.57 & \textbf{67.09} \\
                          & GNNGuard     & 72.75 & 71.03 & 71.47 & 71.52 & 70.36 & \textbf{67.12} \\
                          & SimPGCN      & 74.50 & 73.02 & 72.97 & 72.11 & 70.21 & \textbf{69.97} \\
                          & Avg.          & 72.96 & 71.36 & 72.06 & 71.12 & 70.44 & \textbf{67.49} \\ \midrule
\multirow{6}{*}{Pubmed}   & GCN          & 86.26 & 84.70 & 85.34 & 82.81 & 82.99 & \textbf{77.24} \\
                          & SGC          & 82.21 & 79.49 & 79.99 & 78.69 & 79.51 & \textbf{74.20} \\
                          & GAT          & 85.57 & 83.90 & 84.60 & 83.19 & 82.80 & \textbf{77.83} \\
                          & GNNGuard     & 86.73 & 85.43 & 85.72 & 83.21 & 83.52 & \textbf{77.18} \\
                          & SimPGCN      & 87.91 & 87.23 & 87.24 & 85.95 & 86.69 & \textbf{80.72} \\
                          & Avg.          & 85.73 & 84.15 & 84.58 & 82.77 & 83.10 & \textbf{77.43} \\                          
\bottomrule
\end{tabular}}
\label{tab_res_poi_5}
\end{table}

\subsubsection{Poisoning Attack}
Then, to investigate the performance of GNNs on node classification tasks under poisoning attack, extensive experiments are conducted with the corresponding results being provided in Table \ref{tab_res_poi_5}. LPGIA can improve the performance by approximately 1.5\% and 3\% compared with the best baseline in Cora and Citeseer, respectively. Furthermore, in Cora\_ML and Pubmed, the improved amount is nearly 5\%. For the poisoning attack, the injection will change the feature mapping learned by GNN\cite{strg2023}. This poses a challenge for LPGIA, as our analysis assumes a fixed feature mapping. As expected, LPGIA achieves a smaller improvement in the poisoning attack compared with the evasion attack. However, LPGIA still outperforms all other methods, demonstrating the effectiveness of leveraging the node prediction information.

\begin{figure}[tbp]
    \centering
    \includegraphics[width=0.99\linewidth]{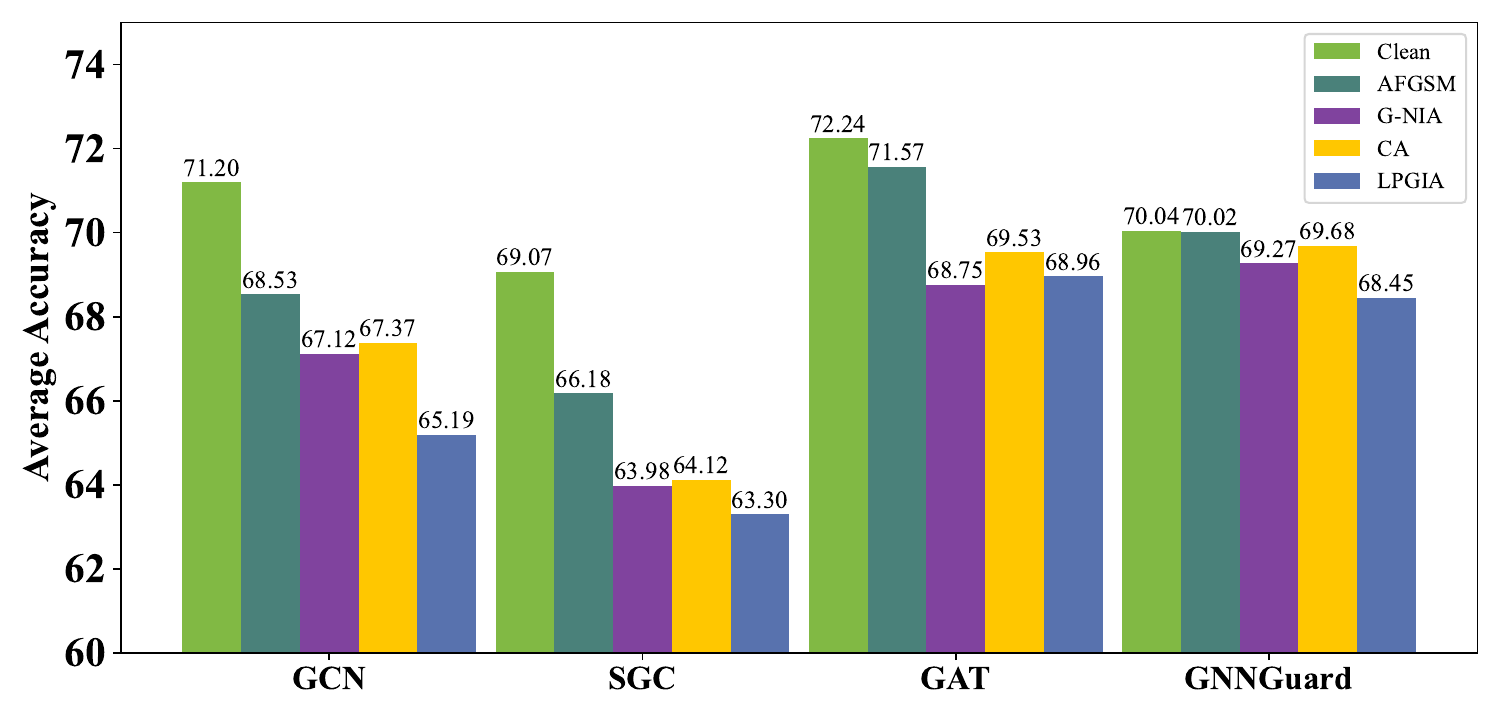}
    \caption{Accuracy (\%) of the GNNs under evasion attack on OGB-Arxiv. Clean indicates the result obtained from the GNN before attacks and CA is short for ClusterAttack.}
    \label{fig_arxiv}
\end{figure}

\subsection{Performance of Attack Methods on Large Datasets}
To evaluate the scalability of LPGIA, we extend our experiments to OGB-Arxiv, a dataset that is significantly larger in scale compared to others. Particularly, OGB-Arxiv stands out as a relatively dense graph, with average node degrees reaching up to 13.7. This implies that the number of potential combinations is larger compared to other datasets in our experimental settings, thereby naturally posing more challenges in terms of time and memory usage. Due to SimPGCN's high space complexity, it is challenging to apply to OGB-Arxiv and will not be included in this experiment. Similarly, due to GANI's high time complexity, our method will be compared with other attack approaches excluding GANI. The evasion attack results on four GNNs by injecting 0.5\% nodes are depicted in Fig. \ref{fig_arxiv}. As the scale of the data increases, a decline in performance is observed across most of the target models, especially SGC, even with a small attack budget. Notably, LPGIA stands out with the best attack performance compared to other baselines in most cases, illustrating its strong scalability to attack GNNs in large-scale graphs.

\subsection{Performance with Different Magnitudes of Injection}
\begin{figure*}[htbp]
\centering
    \includegraphics[width=7in]{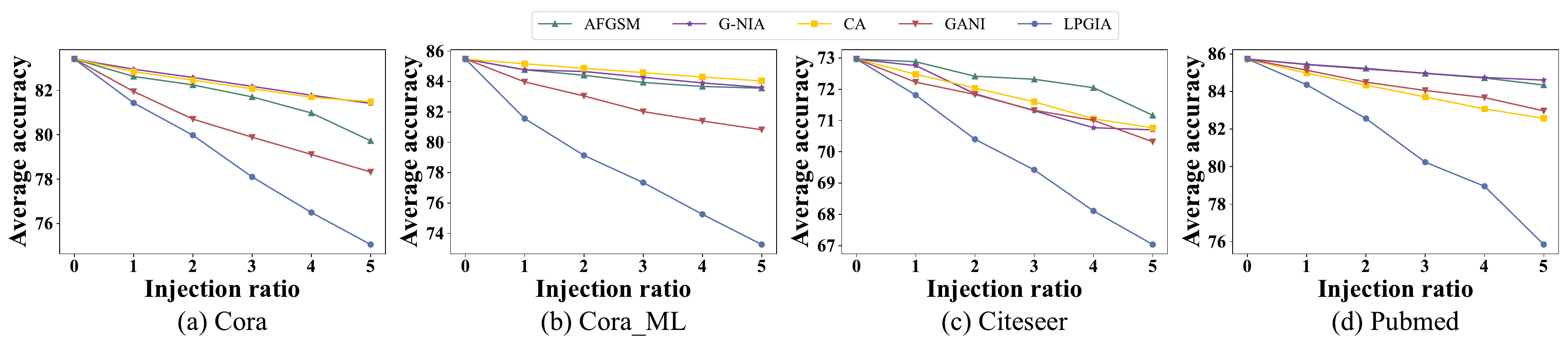}
    \caption{Average accuracy of victim models under evasion attack with injection ratio varying between 1\% and 5\%.}
\label{fig_r15}
\end{figure*}

To reveal the effect of different injection ratios on attack performance, we vary the corresponding value from 1\% to 5\% with the results being illustrated in Fig. \ref{fig_r15}. As indicated, with the increase of injection ratio, the classification accuracy of victim model reduces rapidly. Furthermore, for scenarios with different injection ratios, LPGIA is always of the best attack performance compared with the other considered ones. In the scenario with a 1\% injection ratio, LPGIA shows limited superior attack performance, which is due to the small number of injected nodes. As the injection ratio increases, the superiority of LPGIA becomes more pronounced, suggesting that with a larger number of injected nodes, our method can effectively leverage the similarity of target labels between nodes to perturb the graph, leading to more accuracy reduction of GNN.

\subsection{Performance with Different Hyper-parameters}
\begin{figure*}[htbp]
\centering
    \includegraphics[width=7in]{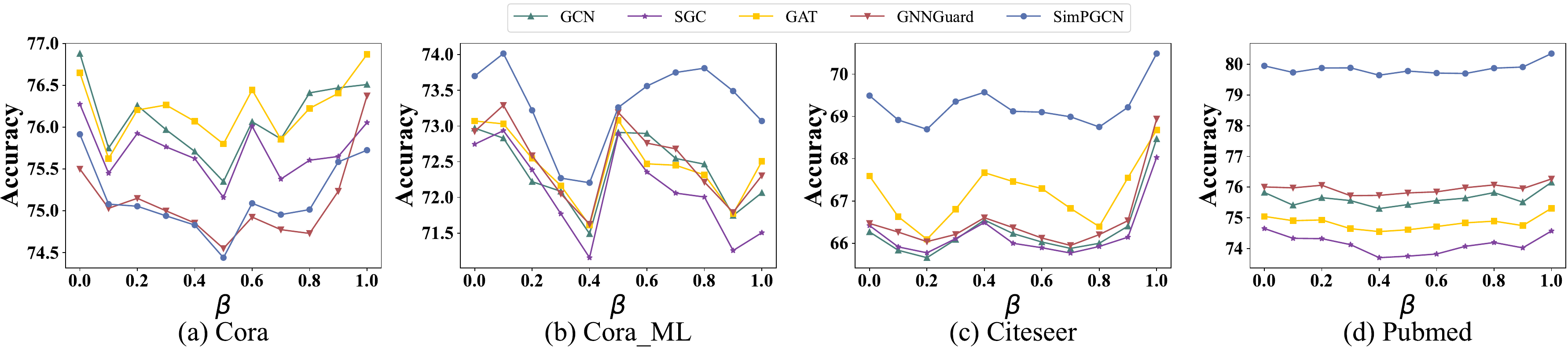}
    \caption{Accuracy of victim models under evasion attack with different $\beta$ in LPGIA. $\beta=1$ means that LPGIA only considers the decreasing extent calculated in Eq. (\ref{eq_s1}). In contrast, $\beta=0$ means that LPGIA only considers the difference calculated in Eq. (\ref{eq_s2}).}
\label{fig_beta}
\end{figure*}

According to Eq. (\ref{eq_s_h}), the propagation score involves a trade-off between two factors, which can be adjusted through the hyper-parameter $\beta$. To investigate the effect of varying $\beta$ on the attack performance, experiments are performed with the results depicted in Fig. \ref{fig_beta}. Note that the other experimental settings remain as those employed in evasion attacks.

As in Fig. \ref{fig_beta}, on Cora\_ML and Pubmed, the optimal $\beta$ is 0.4. Nevertheless, for Cora and Citeseer, the optimal $\beta$ is obtained as 0.5 and 0.2, respectively. It suggests that considering both factors in Eq. (\ref{eq_s_h}) comprehensively is an effective approach, as LPGIA achieves a smaller improvement when relying solely on a single factor for the attack strategy. Furthermore, we observe that LPGIA is sensitive to $\beta$ in most cases. As $\beta$ increases, the obtained accuracy of GNN fluctuates. However, we notice that, in Pubmed, the corresponding accuracy is not sensitive to the varying of hyper-parameter $\beta$. An explanation is that there is not a significant difference between the rankings of nodes based on either of these two factors.

\subsection{Ablation Study}
To better understand the impact of adopting different attack strategies on the performance of LPGIA, we conduct detailed ablation studies. The remaining experimental settings are consistent with those employed in our evasion attack.

\subsubsection{Node Selection Strategy}
First, we conduct experiments to investigate different node selection strategies, and the results are provided in Table \ref{tab_ab_node}. Among them, the ``Random'' strategy indicates that we uniformly assign node scores between $[0, 1]$ at random; the ``Degree'' strategy gives the preference of selecting nodes with low degrees \cite{tdgia2021}. As to the ``HD'' strategy, it only considers the decrease of homophily \cite{gani2022}. As revealed, for the considered binary datasets, i.e., Cora and Citeseer, our node selection strategy is of the best performance. However, our strategy exhibits poor performance than others when considering Cora\_ML, due to the reason that LPGIA selects more nodes with higher degrees than other strategies. Those nodes are hard to attack in Cora\_ML and make our strategy perform worse. Generally speaking, our method is of better capability to identify the valuable target nodes compared with the others as both factors in Eq. (\ref{eq_s_h}) are considered comprehensively.

\begin{table}[t]
\caption{Average accuracy (\%) of GNN under adversarial attack using different node selection strategies in LPGIA. The best results are boldfaced.}
\centering
\begin{tabular}{ccccc}
\toprule
Strategies                         & Cora           & Cora\_ML       & Citeseer       & Pubmed         \\
\midrule
Clean                           & 83.41          & 85.48          & 72.96          & 85.73          \\
Random                          & 77.22          & 76.38          & 70.40          & 77.72          \\
Degree                          & 76.18          & 73.10          & 67.96          & 76.95          \\
HD                              & 76.02          & \textbf{72.59} & 68.24          & 76.42          \\
Ours                            & \textbf{75.06} & 73.27          & \textbf{67.04} & \textbf{75.85} \\
\bottomrule
\end{tabular}
\label{tab_ab_node}
\end{table}

\subsubsection{Cluster Generation Strategy}
As discussed above, the presence of category imbalance within the target label group hinders our ability to mislead the label prediction. Hence, we assess the efficacy of our proposed strategy in mitigating this issue and enhancing attack performance. For comparison, the ``Random'' strategy indicates that we randomly select several nodes in the target label group to form a cluster; The ``TopNodes'' strategy selects the top-$k$ nodes in the target label group with the largest propagation scores to form a cluster. Compared with the ``TopNodes'' strategy, the results demonstrate the effectiveness of the proposed strategy, as shown in Table \ref{tab_ab_cluster}. By alleviating the issue of imbalance, LPGIA achieves improved performance in all cases.

\begin{table}[tb]
\caption{Average accuracy (\%) of GNN under adversarial attack using different cluster generation strategies in LPGIA. The best results are boldfaced.}
\centering
\begin{tabular}{ccccc}
\toprule
Strategies & Cora           & Cora\_ML       & Citeseer       & Pubmed         \\
\midrule
Clean                    & 83.41          & 85.48          & 72.96          & 85.73          \\
Random                   & 78.16          & 76.97          & 70.28          & 77.91          \\
TopNodes                & 76.55          & 73.66          & 68.24          & 76.78          \\
Ours                     & \textbf{75.06} & \textbf{73.27} & \textbf{67.04} & \textbf{75.85} \\
\bottomrule
\end{tabular}
\label{tab_ab_cluster}
\end{table}
\subsubsection{Feature Generation Strategy}
In Table \ref{tab_ab_feature}, we present a comparison of different feature generation strategies. The ``Random'' strategy indicates that we will randomly copy the feature from the original node to the injected node. The ``MostFrequency'' strategy represents a statistics-based method that selects the most frequently occurring feature element among nodes with a specific label. We do not compare with the gradient-based method \cite{tdgia2021} due to the concern about feature inconsistency. The results indicate that the malicious features generated by LPGIA significantly impact the predictions of nodes, compared with other strategies. 

\begin{table}[tb]
\caption{Average accuracy (\%) of GNN under adversarial attack using different feature generation strategies in LPGIA. The best results are boldfaced.}
\centering
\begin{tabular}{ccccc}
\toprule
Strategies & Cora           & Cora\_ML       & Citeseer       & Pubmed         \\
\midrule
Clean                      & 83.41          & 85.48          & 72.96          & 85.73          \\
Random                     & 80.83          & 82.74          & 71.90          & 84.01          \\
MostFrequency              & 78.10          & 78.43          & 70.24          & 82.60          \\
Ours                       & \textbf{75.06} & \textbf{73.27} & \textbf{67.04} & \textbf{75.85} \\
\bottomrule
\end{tabular}
\label{tab_ab_feature}
\end{table}
\subsubsection{Single Module}
\begin{figure}[tbp]
    \centering
    \includegraphics[width=3in]{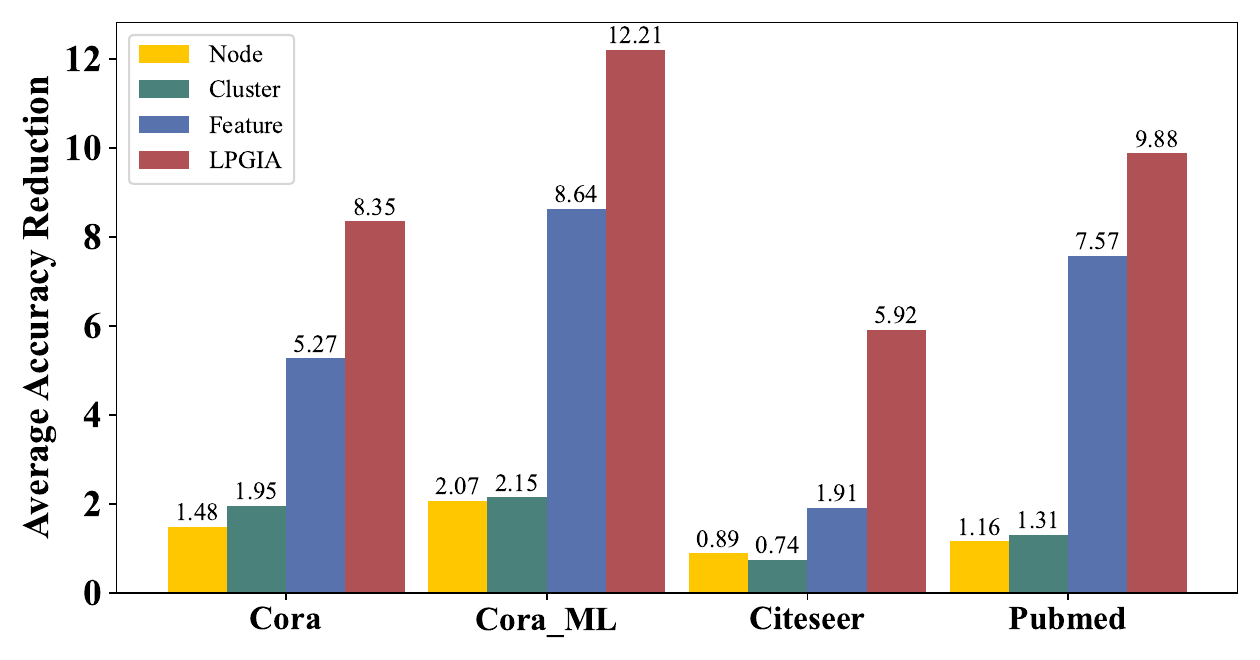}
    \caption{Average accuracy reduction (\%) of GNN under adversarial attack using a single module in LPGIA.}
    \label{fig_smab}
\end{figure}

Finally, Fig. \ref{fig_smab} presents the effectiveness of LPGIA when only one of the proposed strategies is retained and the ``Random'' strategy is applied to the rest module of LPGIA. Note that the feature generation module plays an important role in attacking GNN. A reason for the promising performance of the feature generation module is that we retained the operation of grouping nodes based on the target label. It allows the feature generation module to focus on increasing the probability of the target label for the victim node. However, propagating random features cannot effectively increase the probability of the target label for the victim node, which explains the poor attack performance when using the connection optimization module of LPGIA alone.

Overall, the strategies employed in each module effectively address the challenges. The feature generation module is the most critical component in the LPGIA framework, but it does not diminish the importance of connection modules. Proper combinations of victim nodes can increase the success rate of the attack and facilitate the propagation of malicious features to a greater extent. For LPGIA, the combination of these strategies effectively reduces the accuracy of the GNN model.

\section{Conclusion}\label{sec_con}
In this paper, we study the adversarial attack on homophily graphs and provide a label-propagation-based global injection attack method on GNNs. Specifically, neighboring nodes exhibit significant similarity of their target labels in the homophily graphs, thereby providing valuable insights to us regarding potential directions for attacks. To exploit the relationship between node prediction and graph structure, our method clusters valuable and vulnerable nodes in the graph according to the expected effects of label propagation. Besides, a fake node with a malicious feature will connect to each cluster, increasing the predicted probability of the target label for nodes in the cluster. The extensive experiments on representative datasets demonstrate the superior performance of LPGIA in attacking various GNNs under multiple settings compared to other adversarial attacks.

To improve time efficiency, our work simplifies the propagation process by propagating the smooth probability distribution between nodes, thereby constraining a more accurate propagation simulation. In future work, we intend to deeply explore various propagation processes, especially efficient propagation in the large-scale graph.


\bibliographystyle{IEEEtran}
\bibliography{IEEEabrv,refs.bib}

\begin{thebibliography}{10}
\providecommand{\url}[1]{#1}
\csname url@samestyle\endcsname
\providecommand{\newblock}{\relax}
\providecommand{\bibinfo}[2]{#2}
\providecommand{\BIBentrySTDinterwordspacing}{\spaceskip=0pt\relax}
\providecommand{\BIBentryALTinterwordstretchfactor}{4}
\providecommand{\BIBentryALTinterwordspacing}{\spaceskip=\fontdimen2\font plus
\BIBentryALTinterwordstretchfactor\fontdimen3\font minus \fontdimen4\font\relax}
\providecommand{\BIBforeignlanguage}[2]{{%
\expandafter\ifx\csname l@#1\endcsname\relax
\typeout{** WARNING: IEEEtran.bst: No hyphenation pattern has been}%
\typeout{** loaded for the language `#1'. Using the pattern for}%
\typeout{** the default language instead.}%
\else
\language=\csname l@#1\endcsname
\fi
#2}}
\providecommand{\BIBdecl}{\relax}
\BIBdecl

\bibitem{gcn2017}
T.~N. Kipf and M.~Welling, ``Semi-supervised classification with graph convolutional networks,'' in \emph{Proc. 5th Int. Conf. Learn. Representations}, 2017.

\bibitem{gat2018}
P.~Veličković, G.~Cucurull, A.~Casanova, A.~Romero, P.~Liò, and Y.~Bengio, ``Graph attention networks,'' in \emph{Proc. 6th Int. Conf. Learn. Representations}, 2018.

\bibitem{fagnn2022}
J.~Liu, J.~Zheng, J.~Wu, and Z.~Zheng, ``Fa-gnn: Filter and augment graph neural networks for account classification in ethereum,'' \emph{IEEE Trans. Network Sci. Eng.}, vol.~9, no.~4, pp. 2579--2588, 2022.

\bibitem{ginsd2024}
L.~Cheng, P.~Zhu, K.~Tang, C.~Gao, and Z.~Wang, ``Gin-sd: Source detection in graphs with incomplete nodes via positional encoding and attentive fusion,'' in \emph{Proc. 38th AAAI Conf. Artif. Intell.}, 2024, pp. 55--63.

\bibitem{kyc2024}
D.~Lin, J.~Wu, T.~Huang, K.~Lin, and Z.~Zheng, ``Who is who on ethereum? account labeling using heterophilic graph convolutional network,'' \emph{IEEE Trans. Syst. Man Cybern.: Syst.}, vol.~54, no.~5, pp. 1541--1553, 2024.

\bibitem{sage2017}
W.~L. Hamilton, R.~Ying, and J.~Leskovec, ``Inductive representation learning on large graphs,'' in \emph{Proc. 30th Adv. Neural Inf. Process. Syst.}, 2017, pp. 1025--1035.

\bibitem{compgcn2020}
S.~Vashishth, S.~Sanyal, V.~Nitin, and P.~Talukdar, ``Composition-based multi-relational graph convolutional networks,'' in \emph{Proc. 8th Int. Conf. Learn. Representations}, 2020.

\bibitem{dgcn2023}
C.~Gao, J.~Zhu, F.~Zhang, Z.~Wang, and X.~Li, ``A novel representation learning for dynamic graphs based on graph convolutional networks,'' \emph{IEEE Trans. Cybern.}, vol.~53, no.~6, pp. 3599--3612, 2023.

\bibitem{gin2019}
K.~Xu, W.~Hu, J.~Leskovec, and S.~Jegelka, ``How powerful are graph neural networks?'' in \emph{Proc. 7th Int. Conf. Learn. Representations}, 2019.

\bibitem{ssngc2021}
J.~Wang, P.~Chen, B.~Ma, J.~Zhou, Z.~Ruan, G.~Chen, and Q.~Xuan, ``Sampling subgraph network with application to graph classification,'' \emph{IEEE Trans. Network Sci. Eng.}, vol.~8, no.~4, pp. 3478--3490, 2021.

\bibitem{ahgrr2023}
L.~Sang, M.~Xu, S.~Qian, and X.~Wu, ``Adversarial heterogeneous graph neural network for robust recommendation,'' \emph{IEEE Trans. Comput. Social Syst.}, vol.~10, no.~5, pp. 2660--2671, 2023.

\bibitem{ecpe2024}
P.~Zhu, B.~Wang, K.~Tang, H.~Zhang, X.~Cui, and Z.~Wang, ``A knowledge-guided graph attention network for emotion-cause pair extraction,'' \emph{Knowledge-Based Syst.}, vol. 286, 2024, {A}rt. no. 111342.

\bibitem{uav2022}
J.~Tian, B.~Wang, R.~Guo, Z.~Wang, K.~Cao, and X.~Wang, ``Adversarial attacks and defenses for deep-learning-based unmanned aerial vehicles,'' \emph{IEEE Internet Things J.}, vol.~9, no.~22, pp. 22\,399--22\,409, 2022.

\bibitem{ham2024}
P.~Zhu, Z.~Fan, S.~Guo, K.~Tang, and X.~Li, ``Improving adversarial transferability through hybrid augmentation,'' \emph{Comput. Secur.}, vol. 139, 2024, {A}rt. no. 103674.

\bibitem{lesson2024}
J.~Tian, C.~Shen, B.~Wang, X.~Xia, M.~Zhang, C.~Lin, and Q.~Li, ``Lesson: Multi-label adversarial false data injection attack for deep learning locational detection,'' \emph{IEEE Trans. Dependable Secure Comput., early access}, pp. 1--15, 2024.

\bibitem{gafsi2024}
P.~Zhu, Z.~Pan, Y.~Liu, J.~Tian, K.~Tang, and Z.~Wang, ``A general black-box adversarial attack on graph-based fake news detectors,'' 2024, \emph{arXiv:2404.15744}.

\bibitem{nettack2018}
D.~Z\"{u}gner, A.~Akbarnejad, and S.~G\"{u}nnemann, ``Adversarial attacks on neural networks for graph data,'' in \emph{Proc. 24th ACM SIGKDD Int. Conf. Knowl. Discov. Data Mining}, 2018, pp. 2847--2856.

\bibitem{rls2v2018}
H.~Dai, H.~Li, T.~Tian, X.~Huang, L.~Wang, J.~Zhu, and L.~Song, ``Adversarial attack on graph structured data,'' in \emph{Proc. Int. Conf. Mach. Learn.}, 2018, pp. 1115--1124.

\bibitem{mettack2019}
D.~Zügner and S.~Günnemann, ``Adversarial attacks on graph neural networks via meta learning,'' in \emph{Proc. 7th Int. Conf. Learn. Representations}, 2019.

\bibitem{pgd2019}
K.~Xu, H.~Chen, S.~Liu, P.-Y. Chen, T.-W. Weng, M.~Hong, and X.~Lin, ``Topology attack and defense for graph neural networks: An optimization perspective,'' in \emph{Proc. 28th Int. Joint Conf. Artif. Intell.}, 2019, pp. 3961--3967.

\bibitem{linkatk2020}
J.~Chen, X.~Lin, Z.~Shi, and Y.~Liu, ``Link prediction adversarial attack via iterative gradient attack,'' \emph{IEEE Trans. Comput. Social Syst.}, vol.~7, no.~4, pp. 1081--1094, 2020.

\bibitem{sga2021}
J.~Li, T.~Xie, L.~Chen, F.~Xie, X.~He, and Z.~Zheng, ``Adversarial attack on large scale graph,'' \emph{IEEE Trans. Knowl. Data Eng.}, vol.~35, no.~1, pp. 82--95, 2023.

\bibitem{graphatker2022}
J.~Chen, D.~Zhang, Z.~Ming, K.~Huang, W.~Jiang, and C.~Cui, ``Graphattacker: A general multi-task graph attack framework,'' \emph{IEEE Trans. Network Sci. Eng.}, vol.~9, no.~2, pp. 577--595, 2022.

\bibitem{gf2023}
J.~Chen, G.~Huang, H.~Zheng, S.~Yu, W.~Jiang, and C.~Cui, ``Graph-fraudster: Adversarial attacks on graph neural network-based vertical federated learning,'' \emph{IEEE Trans. Comput. Social Syst.}, vol.~10, no.~2, pp. 492--506, 2023.

\bibitem{nipa2020}
Y.~Sun, S.~Wang, X.~Tang, T.-Y. Hsieh, and V.~Honavar, ``Adversarial attacks on graph neural networks via node injections: A hierarchical reinforcement learning approach,'' in \emph{Proc. ACM World Wide Web Conf.}, 2020, pp. 673--683.

\bibitem{afgsm2020}
J.~Wang, M.~Luo, F.~Suya, J.~Li, Z.~Yang, and Q.~Zheng, ``Scalable attack on graph data by injecting vicious nodes,'' \emph{Data Min. Knowl. Discovery}, vol.~34, no.~5, pp. 1363--1389, 2020.

\bibitem{tdgia2021}
X.~Zou, Q.~Zheng, Y.~Dong, X.~Guan, E.~Kharlamov, J.~Lu, and J.~Tang, ``Tdgia: Effective injection attacks on graph neural networks,'' in \emph{Proc. 27th ACM SIGKDD Int. Conf. Knowl. Discov. Data Mining}, 2021, pp. 2461--2471.

\bibitem{gnia2021}
S.~Tao, Q.~Cao, H.~Shen, J.~Huang, Y.~Wu, and X.~Cheng, ``Single node injection attack against graph neural networks,'' in \emph{Proc. 30th ACM Int. Conf. Knowl. Manage.}, 2021, pp. 1794--1803.

\bibitem{ca2022}
Z.~Wang, Z.~Hao, Z.~Wang, H.~Su, and J.~Zhu, ``Cluster attack: Query-based adversarial attacks on graph with graph-dependent priors,'' in \emph{Proc. 31st Int. Joint Conf. Artif. Intell.}, 2022, pp. 768--775.

\bibitem{appna2019}
J.~Gasteiger, A.~Bojchevski, and S.~Günnemann, ``Combining neural networks with personalized pagerank for classification on graphs,'' in \emph{Proc. 7th Int. Conf. Learn. Representations}, 2019.

\bibitem{gcnlpa2021}
H.~Wang and J.~Leskovec, ``Combining graph convolutional neural networks and label propagation,'' \emph{ACM Trans. Inf. Syst.}, vol.~40, no.~4, pp. 1--27, 2021.

\bibitem{pta2021}
H.~Dong, J.~Chen, F.~Feng, X.~He, S.~Bi, Z.~Ding, and P.~Cui, ``On the equivalence of decoupled graph convolution network and label propagation,'' in \emph{Proc. ACM World Wide Web Conf.}, 2021, pp. 3651--3662.

\bibitem{cs2021}
Q.~Huang, H.~He, A.~Singh, S.-N. Lim, and A.~Benson, ``Combining label propagation and simple models out-performs graph neural networks,'' in \emph{Proc. 9th Int. Conf. Learn. Representations}, 2021.

\bibitem{lp2002}
X.~Zhu and Z.~Ghahramani, ``Learning from labeled and unlabeled data with label propagation,'' \emph{Technical Report, Carnegie Mellon University}, 2002.

\bibitem{llg2003}
D.~Zhou, O.~Bousquet, T.~Lal, J.~Weston, and B.~Sch\"{o}lkopf, ``Learning with local and global consistency,'' in \emph{Proc. 16th Adv. Neural Inf. Process. Syst.}, 2003, pp. 321--328.

\bibitem{lpch2007}
U.~N. Raghavan, R.~Albert, and S.~Kumara, ``Near linear time algorithm to detect community structures in large-scale networks,'' \emph{Phys. Rev. E}, vol.~76, no.~3, 2007, {A}rt. no. 036106.

\bibitem{lsa2022}
H.~Wang, Y.~Liu, P.~Yin, H.~Zhang, X.~Xu, and Q.~Wen, ``Label specificity attack: Change your label as i want,'' \emph{Int. J. Intell. Syst.}, vol.~37, no.~10, pp. 7767--7786, 2022.

\bibitem{g2snia2023}
D.~Chen, J.~Zhang, Y.~Lv, J.~Wang, H.~Ni, S.~Yu, Z.~Wang, and Q.~Xuan, ``Single node injection label specificity attack on graph neural networks via reinforcement learning,'' 2023, \emph{arXiv:2305.02901}.

\bibitem{mga2021}
J.~Chen, Y.~Chen, H.~Zheng, S.~Shen, S.~Yu, D.~Zhang, and Q.~Xuan, ``Mga: Momentum gradient attack on network,'' \emph{IEEE Trans. Comput. Social Syst.}, vol.~8, no.~1, pp. 99--109, 2021.

\bibitem{epoatk2023}
X.~Lin, C.~Zhou, J.~Wu, H.~Yang, H.~Wang, Y.~Cao, and B.~Wang, ``Exploratory adversarial attacks on graph neural networks for semi-supervised node classification,'' \emph{Pattern Recognit.}, vol. 133, 2023, {A}rt. no. 109042.

\bibitem{cana2023}
S.~Tao, Q.~Cao, H.~Shen, Y.~Wu, L.~Hou, F.~Sun, and X.~Cheng, ``Adversarial camouflage for node injection attack on graphs,'' \emph{Information Sciences}, vol. 649, 2023, {A}rt. no. 119611.

\bibitem{hao2022}
Y.~Chen, H.~Yang, Y.~Zhang, M.~KAILI, T.~Liu, B.~Han, and J.~Cheng, ``Understanding and improving graph injection attack by promoting unnoticeability,'' in \emph{Proc. 10th Int. Conf. Learn. Representations}, 2022.

\bibitem{gani2022}
J.~Fang, H.~Wen, J.~Wu, Q.~Xuan, Z.~Zheng, and C.~K. Tse, ``Gani: Global attacks on graph neural networks via imperceptible node injections,'' \emph{IEEE Trans. Comput. Social Syst., early access}, pp. 1--14, 2024.

\bibitem{g2a2c2023}
M.~Ju, Y.~Fan, C.~Zhang, and Y.~Ye, ``Let graph be the go board: gradient-free node injection attack for graph neural networks via reinforcement learning,'' in \emph{Proc. 37th AAAI Conf. Artif. Intell.}, 2023, pp. 4383--4390.

\bibitem{geom2020}
P.~Hongbin, W.~Bingzhe, C.~Kevin Chen-Chuan, L.~Yu, and Y.~Bo, ``Geom-gcn: Geometric graph convolutional networks,'' in \emph{Proc. 8th Int. Conf. Learn. Representations}, 2020.

\bibitem{h2gcn2020}
J.~Zhu, Y.~Yan, L.~Zhao, M.~Heimann, L.~Akoglu, and D.~Koutra, ``Beyond homophily in graph neural networks: Current limitations and effective designs,'' in \emph{Proc. 33th Adv. Neural Inf. Process. Syst.}, 2020, pp. 7793--7804.

\bibitem{planetoid2016}
Z.~Yang, W.~Cohen, and R.~Salakhudinov, ``Revisiting semi-supervised learning with graph embeddings,'' in \emph{Proc. 33rd Int. Conf. Mach. Learn.}, 2016, pp. 40--48.

\bibitem{coraml2018}
A.~Bojchevski and S.~Günnemann, ``Deep gaussian embedding of graphs: Unsupervised inductive learning via ranking,'' in \emph{Proc. 6th Int. Conf. Learn. Representations}, 2018.

\bibitem{ogb2020}
W.~Hu, M.~Fey, M.~Zitnik, Y.~Dong, H.~Ren, B.~Liu, M.~Catasta, and J.~Leskovec, ``Open graph benchmark: Datasets for machine learning on graphs,'' in \emph{Proc. 33th Adv. Neural Inf. Process. Syst.}, 2020, pp. 22\,118--22\,133.

\bibitem{sgc2019}
F.~Wu, A.~Souza, T.~Zhang, C.~Fifty, T.~Yu, and K.~Weinberger, ``Simplifying graph convolutional networks,'' in \emph{Proc. 36th Int. Conf. Mach. Learn.}, vol.~97, 2019, pp. 6861--6871.

\bibitem{gnnguard2020}
X.~Zhang and M.~Zitnik, ``Gnnguard: Defending graph neural networks against adversarial attacks,'' in \emph{Proc. 33rd Adv. Neural Inf. Process. Syst.}, vol.~33, 2020, pp. 9263--9275.

\bibitem{simpgcn2021}
W.~Jin, T.~Derr, Y.~Wang, Y.~Ma, Z.~Liu, and J.~Tang, ``Node similarity preserving graph convolutional networks,'' in \emph{Proc. 14th ACM Int. Conf. Web Search Data Mining}, 2021, pp. 148--156.

\bibitem{strg2023}
K.~Li, Y.~Liu, X.~Ao, and Q.~He, ``Revisiting graph adversarial attack and defense from a data distribution perspective,'' in \emph{Proc. 11th Int. Conf. Learn. Representations}, 2023.

\end{thebibliography}

\begin{IEEEbiography} [{\includegraphics[width=0.9in,height=1.25in,clip,keepaspectratio]{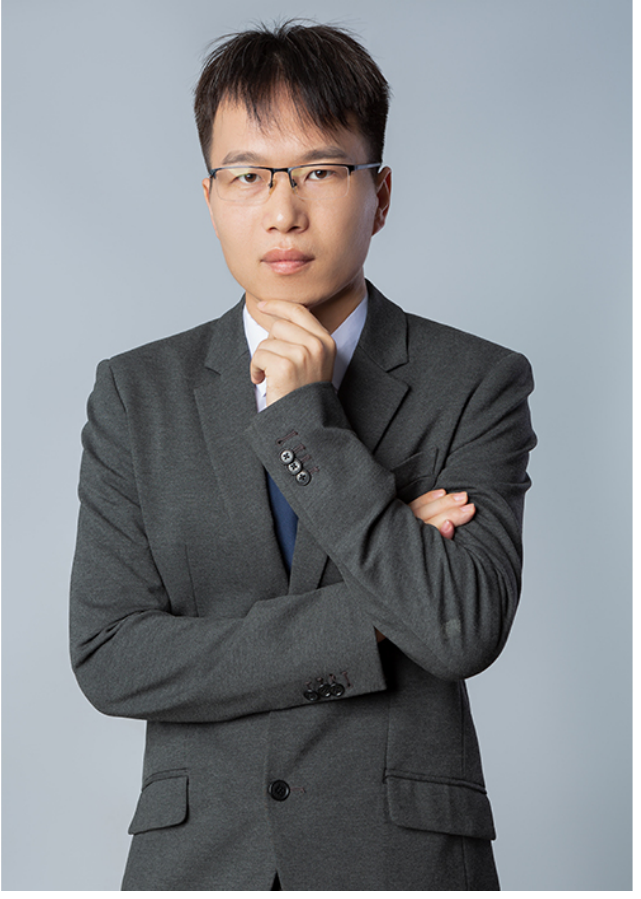}}] {Peican Zhu} (Member, IEEE) received the Ph.D. degree from the University of Alberta, Edmonton, AB, Canada, in 2015. He is currently an Associate Professor with the School of Artificial Intelligence, Optics and Electronics (iOPEN), Northwestern Polytechnical University, Xi’an, China. His research interests include data-driven complex systems modeling, complex social network analysis, and AI security.
\end{IEEEbiography}

\begin{IEEEbiography} [{\includegraphics[width=0.9in,height=1.25in,clip,keepaspectratio]{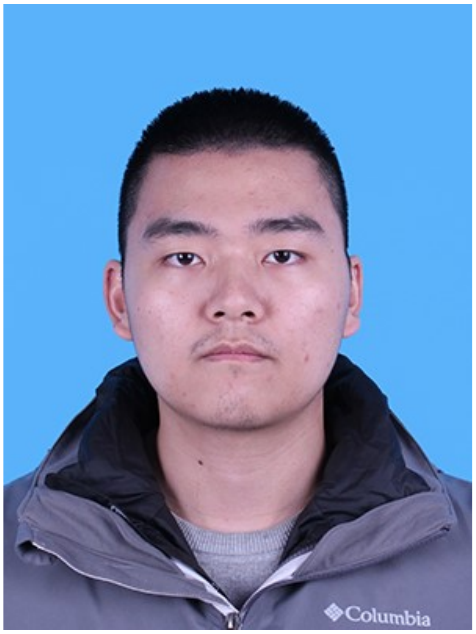}}] {Zechen Pan} received the bachelor's degree from Northwestern Polytechnical University, Xi’an, China, in 2022. He is currently working toward the M.S. degree in the School of Computer Science, Northwestern Polytechnical University. His research interests include complex social network analysis and AI security.
\end{IEEEbiography}

\begin{IEEEbiography} [{\includegraphics[width=0.9in,height=1.25in,clip,keepaspectratio]{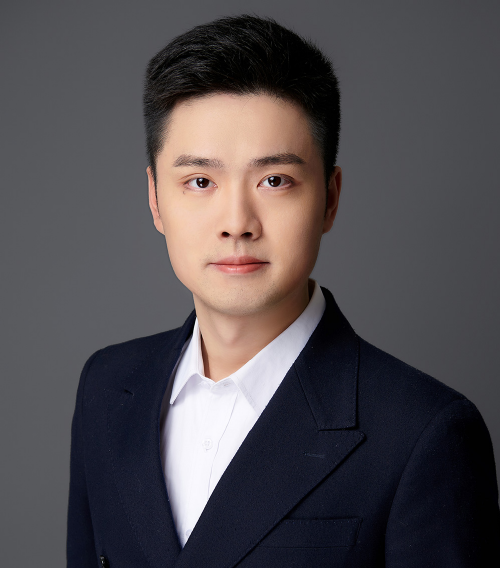}}] {Keke Tang} (Member, IEEE) received the Ph.D. degree from the University of Science and Technology of China, Hefei, China, in 2017. He is currently an Associate Professor with Guangzhou University, Guangzhou, China. Prior to joining Guangzhou University in 2019, he was a Postdoctoral Fellow with the University of Hong Kong, Hong Kong, China. His research interests fall into the areas of robotics, computer vision, computer graphics, and cyberspace security.
\end{IEEEbiography}

\begin{IEEEbiography} [{\includegraphics[width=0.9in,height=1.25in,clip,keepaspectratio]{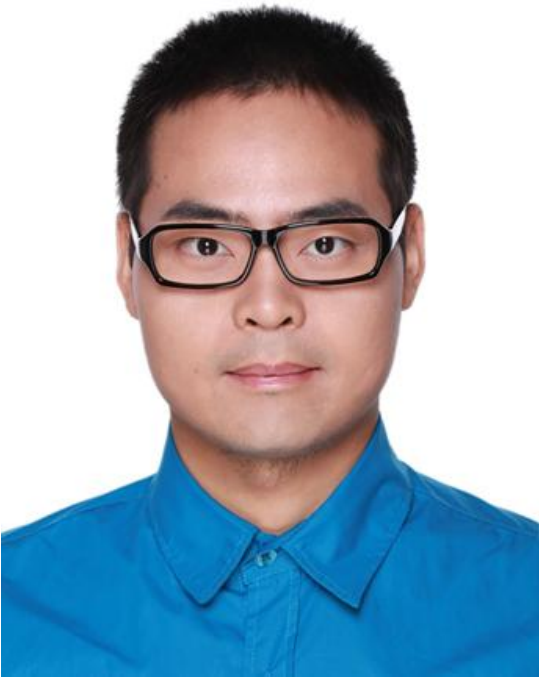}}] {Xiaodong Cui} received the Ph.D. degree in electrical engineering from the University of Texas at San Antonio in 2015. Currently, he is an Associate Professor at the College of Marine Science and Technology in Northwestern Polytechnical University, Xi'an, China. His research interests include underwater acoustic target recognition, object detection and dereverberation, AI security.
\end{IEEEbiography}

\begin{IEEEbiography} [{\includegraphics[width=0.9in,height=1.25in,clip,keepaspectratio]{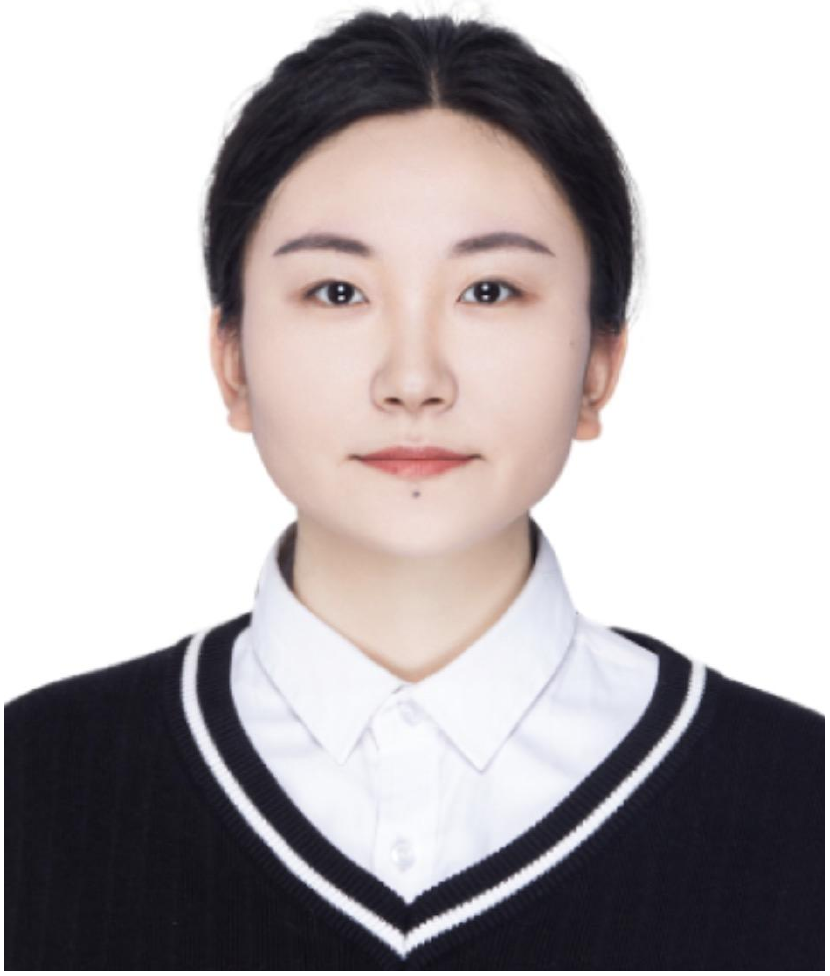}}] {Jinhuan Wang} received the B.S. in automation and M.S. degrees in control science and engineering with the College of Information Engineering, Zhejiang University of Technology, Hangzhou, China, in 2017 and 2020, respectively, where she is currently working toward the Ph.D. degree in control science and engineering. Her research interests include social network data
mining and machine learning.
\end{IEEEbiography}

\begin{IEEEbiography} [{\includegraphics[width=0.9in,height=1.25in,clip,keepaspectratio]{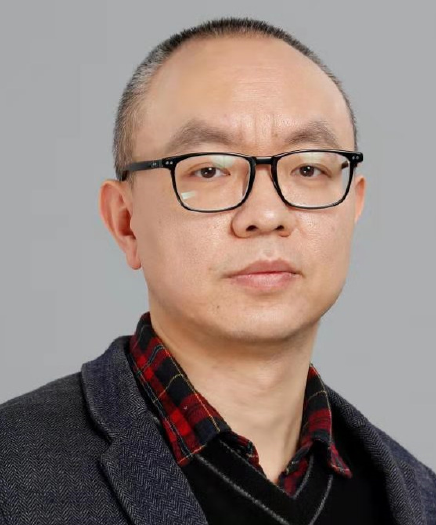}}] {Qi Xuan} (Senior Member, IEEE) received the B.S. and Ph.D. degrees in control theory and engineering from Zhejiang University, Hangzhou, China, in 2003 and 2008, respectively. He was a Post-Doctoral Researcher with the Department of Information Science and Electronic Engineering, Zhejiang University, from 2008 to 2010, respectively, and a Research Assistant with the Department of Electronic Engineering, City University of Hong Kong, Hong Kong, in 2010 and 2017. From 2012 to 2014, he was a Post-Doctoral Fellow with the Department of Computer Science, University of California at Davis, CA, USA. He is a senior member of the IEEE and is currently a Professor with the Institute of Cyberspace Security, College of Information Engineering, Zhejiang University of Technology, Hangzhou, China. His current research interests include network science, graph data mining, cyberspace security, machine learning, and computer vision.
\end{IEEEbiography}

\end{document}